\documentclass[acmtog, nonacm]{acmart}
\AtBeginDocument{%
  \providecommand\BibTeX{{%
    \normalfont B\kern-0.5em{\scshape i\kern-0.25em b}\kern-0.8em\TeX}}}

\copyrightyear{2023}
\setcopyright{rightsretained}
\acmConference[SA Conference Papers '23]{SIGGRAPH Asia 2023 Conference Papers}{December 12--15, 2023}{Sydney, NSW, Australia}
\acmBooktitle{SIGGRAPH Asia 2023 Conference Papers (SA Conference Papers '23), December 12--15, 2023, Sydney, NSW, Australia}
\acmDOI{10.1145/3610548.3618250}
\acmISBN{979-8-4007-0315-7/23/12}

\usepackage{xcolor}
\usepackage{algorithm}
\usepackage[noend]{algpseudocode}
\usepackage{siunitx}
\newcommand{\revised}[1]{{#1}}

\newcommand{\revisedb}[1]{\textcolor{black}{{#1}}}

\DeclareMathOperator*{\argmin}{argmin}

\begin{document}

\title{Simultaneous Color Computer Generated Holography}

\author{Eric Markley}
\email{emarkley@berkeley.edu}
\orcid{1234-5678-9012}
\affiliation{%
  \institution{Reality Labs Research, Meta}
  %\streetaddress{9805 Willows Road NE}
  %\city{Redmond}
  %\state{Washington}
  \country{USA}
  %\postcode{98052}
}

\author{Nathan Matsuda}
\email{nathan.matsuda@meta.com}
\affiliation{%
  \institution{Reality Labs Research, Meta}
  %\streetaddress{9805 Willows Road NE}
  %\city{Redmond}
  %\state{Washington}
  \country{USA}
  %\postcode{98052}
  }

\author{Florian Schiffers}
\email{fschiffers@meta.com}
\affiliation{%
  \institution{Reality Labs Research, Meta}
  %\streetaddress{9805 Willows Road NE}
  %\city{Redmond}
  %\state{Washington}
  \country{USA}
  %\postcode{98052}
  }

\author{Oliver Cossairt}

\email{ocossairt@meta.com}
\affiliation{%
  \institution{Reality Labs Research, Meta}
  %\streetaddress{9805 Willows Road NE}
  %\city{Redmond}
  %\state{Washington}
  \country{USA}
  %\postcode{98052}
  }

\author{Grace Kuo}

\email{gracekuo@meta.com}
\affiliation{%
  \institution{Reality Labs Research, Meta}
  %\streetaddress{9805 Willows Road NE}
  %\city{Redmond}
  %\state{Washington}
  \country{USA}
  %\postcode{98052}
  }

\renewcommand{\shortauthors}{Markley, et al.}

\begin{abstract}
Computer generated holography has long been touted as the future of augmented and virtual reality (AR/VR) displays, but has yet to be realized in practice.
 Previous high-quality, color holographic displays have made either a 3$\times$ sacrifice on frame rate by using a sequential color illumination scheme or used more than one spatial light modulator (SLM) and/or bulky, complex optical setups.
The reduced frame rate of sequential color introduces distracting judder and color fringing in the presence of head motion while the form factor of current simultaneous color systems is incompatible with a head-mounted display.
In this work, we propose a framework for simultaneous color holography that allows the use of the full SLM frame rate while maintaining a compact and simple optical setup.
Simultaneous color holograms are optimized through the use of a perceptual loss function, a physics-based neural network wavefront propagator, and a camera-calibrated forward model. 
We measurably improve hologram quality compared to other simultaneous color methods and move one step closer to the realization of color holographic displays for AR/VR.

% We measurably improve hologram quality compared to other simultaneous color holography methods such as bit and depth division and move one step closer to the realization of color holographic displays for AR/VR.

\end{abstract}

\begin{teaserfigure}
  \includegraphics[clip, trim=0in 9.25in 0in 0in, width=\textwidth]{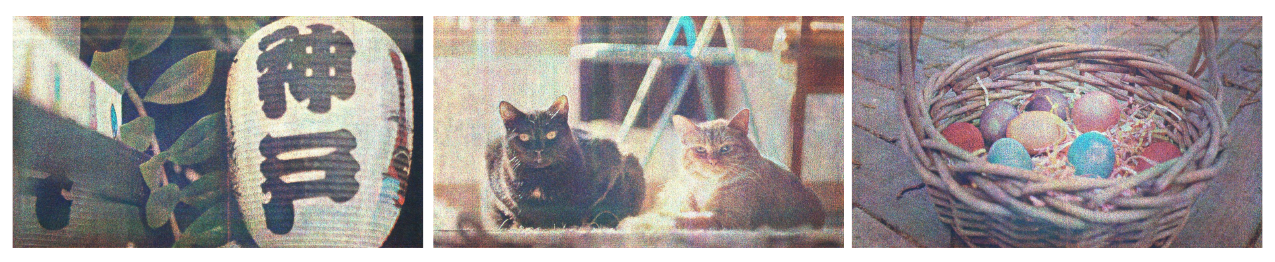}
  \caption{\textbf{Simultaneous color holograms captured in experiment.}  Traditionally, color holograms are illuminated sequentially with a unique spatial light modulator (SLM) pattern for each color channel.  In this work we outline a flexible framework that enables the use of a single SLM pattern for red-green-blue (RGB) holograms using simultaneous RGB illumination.  We validate this framework experimentally on a simple and compact optical setup.
  }
  \Description{Simultaneous color holography experimental setup and results}
  \label{fig:teaser}
\end{teaserfigure}

% \received{25 January 2023}
% \received[revised]{12 March 2023}
% \received[accepted]{5 June 2023}

\maketitle

%%%%%%%%%%%%%%%%%%%%%%%%%%%%%%%%%%%%%%%%%%%
% Main Text

\section{Introduction}
\label{Intro}
Holographic displays are a promising technology for augmented and virtual reality (AR/VR). Such displays use a spatial light modulator (SLM) to shape an incoming coherent wavefront so that it appears as though the wavefront came from a real, three-dimensional (3D) object. The resulting image can have natural defocus cues, providing a path to resolve the vergance-accommodation conflict of stereoscopic displays \cite{kim2022accommodative}.  Additionally, the fine-grain control offered by holography can also correct for optical aberrations, provide custom eyeglass prescription correction in software, and enable compact form-factors \cite{maimone2017holographic}, while improving light efficiency compared to traditional LCD or OLED displays \cite{Yin2022AdvancedApplications}. Recent publications have demonstrated significant improvement in hologram image quality \cite{maimone2017holographic, Peng2020NeuralTraining, Choi2021NeuralDisplays} and computation time \cite{shi2021towards, eybposh2020high}, but color holography for AR/VR has remained an open problem.

Traditionally, red-green-blue (RGB) holograms are created through \textit{field sequential color}, where a separate hologram is computed for each of the three wavelengths; these are displayed in sequence and synchronized with the color of the illumination source. Due to persistence of vision, this appears as a single full color image if the update is sufficiently fast, enabling color holography for static displays. However, in a head-mounted AR/VR system displaying world-locked content, frame rate requirements are higher to prevent noticeable judder \cite{van2016asynchronous}. In fact, all modern VR displays are ``low persistance'' meaning the image content is only displayed for a fraction of the frame time (usually about 10\%) and no content is shown during the rest of the frame \cite{zielinski2015exploring}. This is usually achieved by strobing the illumination, but if one wished to display three sequential color frames all within a 10\% persistence time, it would require the display to update $30\times$ faster than the effective frame rate. Without low persistence, field sequential color leads to strong color fringing (visible spatial separation of the colors) particularly when the user rotates their head while tracking a fixed object with their eyes \cite{riecke2006selected}.

Low frame rate displays exacerbate these artifacts, and the most common SLM technology for holography, liquid-crystal-on-silicon (LCoS), is quite slow due to the physical response time of the liquid crystal (LC) layer \cite{zhang2014fundamentals}. Although most commercial LCoS SLMs can be driven at 60 Hz, at that speed the SLM will have residual artifacts from the prior frames \cite{haist2015holography}. High speed SLMs based on micro-electro-mechanical system (MEMS)~\cite{MEMS, Choi2022Time-multiplexedModulators} or dual-frequency LCoS~\cite{serati2003highresolution} are becoming more widely available, but even with these devices, simultaneous color is desirable since it eliminates color fringing, enables low persistence, and frees temporal bandwidth for other uses, such as increasing the effective etendue by scanning the field of view or eyebox position \cite{lee2020wide}.

% SLMs can be much faster (in the kilohertz range) but so far have larger pixels and limited bit depth~\cite{MEMS, Choi2022Time-multiplexedModulators}.

%Dual-frequency LCoS technology has recently enabled sub-millisecond response time for LCoS SLMs of standard phase range~\cite{serati2003highresolution}.  This technology is compatible with an extended phase range, and the authors hope to see this commercially available in the near future.

In this work, we aim to display RGB holograms using only a single SLM pattern, enabling a $3\times$ increase in frame rate compared to sequential color and completely removing color fringing artifacts. Our compact setup does not use a physical filter in the Fourier plane or bulky optics to combine color channels. Instead, the full SLM is simultaneously illuminated by an on-axis RGB source, and we optimize the SLM pattern to form the full color image. We design a flexible framework for end-to-end optimization of the digital SLM input from the target RGB intensity, allowing us to optimize for SLMs with extended phase range, and we develop a color-specific perceptual loss function which further improves color fidelity. Our method is validated experimentally on 2D and 3D content.
%
% Rather than improve upon existing SLM hardware, this work aims to increase the effective frame rate of current LCOS SLMs for color holography through the use of simultaneous illumination.  To achieve this in a simple and compact form factor, an unfiltered holography setup with on axis illumination is used. The compact setup results in a complex optical forward model involving higher diffraction orders and neural networks requiring calibration.  The calibrated, differentiable forward model allows a gradient based approach to hologram optimization to be taken.  By using a gradient based method, a custom, perceptual loss function can be used helping making the optimization problem better posed.  Finally, camera-based calibration and active camera-in-the-loop are used to achieve state-of-the-art image quality.
  
Specifically, we make the following contributions:

\begin{itemize}
    \item We introduce a novel algorithm for generating simultaneous color holograms which takes advantage of the extended phase range of the SLM in an end-to-end manner and uses a new loss function based on human color perception.
    \item We analyze the ``depth replicas'' artifact in simultaneous color holography and demonstrate how these replicas can be mitigated with extended phase range.
    \item We demonstrate experimental simultaneous color holograms in both 2D and 3D using a custom camera-calibrated model.
\end{itemize}

\begin{figure*}
    \centering
    \includegraphics[clip, trim = 0in 8.5in 0in 0in, width=.99\textwidth]{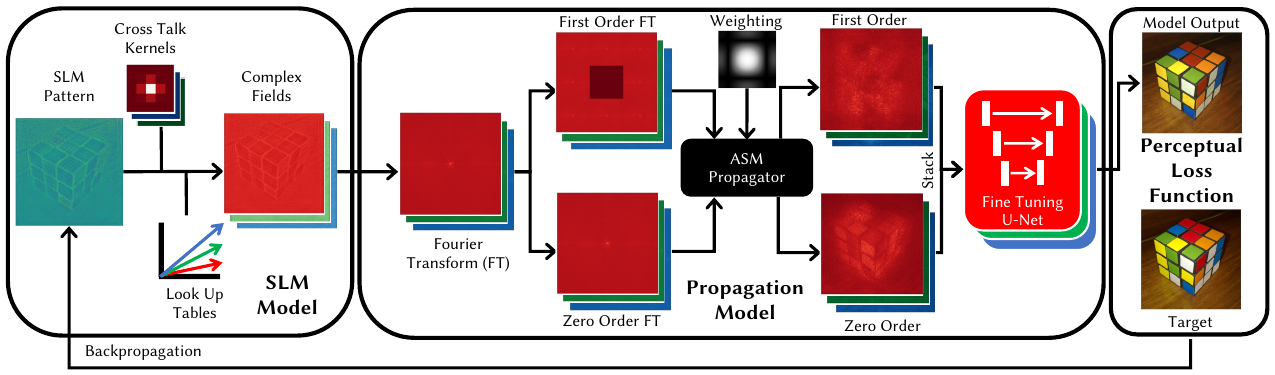}
    \caption{\textbf{Hologram optimization framework.}
    This figure illustrates the three key components of the simultaneous color optimization framework: an SLM model, a propagation model, and a perceptual loss function. The SLM model maps voltage values to a complex field using a learned cross-talk kernel and a linear lookup table. The complex wavefront from the SLM is then propagated to the sensor plane using a modified version of the model proposed by \citet{Gopakumar2021UnfilteredDisplays}, which separates the zeroth and first diffraction orders and combines them through a U-Net. The output is then fed into the perceptual loss function, and gradients are calculated using PyTorch's autograd implementation. The SLM voltages are then updated using these gradients. Rubik's cube source image by Iwan Gabovitch (CC BY 2.0).}  
    \label{fig:ForwardModel}
\end{figure*}

\section{Related Works}
\label{RelatedWorks}
%In this section we discuss previous work in computer generated holography (CGH).  We begin with an overview of current approaches to color holography. These approaches can be split into three main categories: temporal multiplexing, spatial multiplexing, and color multiplexing coding \cite{Pi2022ReviewDisplay}. 
% \fs{I like an overview sentence. But if there's not much space let's leave it out.}
% \paragraph{Temporal Multiplexing Holography}
% \label{TMH}
% \subsection{Color Holography}

\paragraph{Field Sequential Color}
The vast majority of color holographic displays use field sequential color in which the SLM is sequentially illuminated by red, green, and blue sources while the SLM pattern is updated accordingly \cite{maimone2017holographic, jang2018holographic, Chakravarthula2019WirtingerDisplays, Chakravarthula2020LearnedDisplays, chakravarthula2022pupil,  Peng2020NeuralTraining, Peng2021Speckle-freeCalibration,Choi2021NeuralDisplays, choi2021optimizing, shi2021towards, yang2022diffraction, li2016holographic}. Field sequential color is effective at producing full color holograms but reduces frame rate by a factor of $3\times$ and creates color fringing artifacts in the presence of head motion. \revised{These limitations are not alleviated by recent work where the color of each sub-frame is manipulated to increase peak brightness~\cite{kavakli2023holohdr}, and they present a particular challenge for LCoS SLMs where refresh rate is severely limited by the LC response time \cite{zhang2014fundamentals}.} Although, SLMs based on MEMS technology can run at high frame rates in the kilohertz range \cite{MEMS}, so far these modulators are maximum 4-bit displays, with most being binary \cite{Choi2022Time-multiplexedModulators, kim2022accommodative, lee2022high}. Even with emerging 8-bit high frame rate modulators~\cite{serati2003highresolution}, it may be worthwhile to maintain the full temporal bandwidth, since the extra bandwidth can be used to address other holography limitations. For example, speckle can be reduced through temporal averaging \cite{Choi2022Time-multiplexedModulators, kim2022accommodative, lee2022high}, and limited etendue can be mitigated through pupil scanning \cite{jang2018holographic, kim2022holographic}.

\paragraph{Spatial Multiplexing} An alternate approach is spatial multiplexing, which maintains the native SLM frame rate by using different regions of the SLM for each color. Most prior works in this area use three separate SLMs and an array of optics to combine the wavefronts~\cite{Yaras2009Real-timeIllumination, Shiraki2009SimplifiedLinks, Nakayama2010Real-timePanels}. 
Although this method produces high quality holograms, the resulting systems are  bulky, expensive, and require precise alignment, making them poorly suited for near-eye displays. Spatial multiplexing can also be implemented with a single SLM split into sub-regions~\cite{Makowski2011SimpleColor, Makowski2009ExperimentalDisplay}; while less expensive, this approach still requires bulky combining optics and sacrifices space-bandwidth product (SBP), also known as etendue. Etendue is already a limiting factor in holographic displays \cite{kuo2020high}, and further reduction limits the range of viewing angles or display field-of-view.

\paragraph{Frequency Multiplexing} Rather than split the physical extent of the SLM into regions, frequency multiplexing assigns each color a different region in the frequency domain, and the colors are separated with a physical color filter at the Fourier plane of a 4$f$ system \cite{Makowski2010ColorHolograms, Lin17, Lin19}. A variation on this idea uses different angles of illumination for each color so that the physical filter in Fourier space is not color-specific \cite{Xue:14}. Frequency multiplexing can also be implemented with white light illumination, which reduces speckle noise at the cost of resolution \cite{Kozacki16, yang2019full}. However, all of these techniques involve filtering in Fourier space, which sacrifices system etendue and requires a bulky 4$f$ system.

\paragraph{Depth Division and Bit Division for Simultaneous Color} The prior methods most closely related to our work also use simultaneous RGB illumination over the SLM, maintain the full SLM etendue, and don't require a bulky 4$f$ system \cite{Pi2022ReviewDisplay}. We refer to the first method as \textit{depth division multiplexing} which takes advantage of the ambiguity between color and propagation distance (explained in detail in Sec. \ref{sec:color-depth-ambiguity}) and assigns each color a different depth~\cite{Makowski2008ColorfulHologram, Makowski2010ColorHolograms}. After optimizing with a single color for the correct multiplane image, the authors show they can form a full color 2D hologram when illuminating in RGB. However,  this approach does not account for wavelength dependence of the SLM response, and since it explicitly defines content at multiple planes, it translates poorly to 3D.

Another similar approach is \textit{bit division multiplexing}, which takes advantage of the extended phase range of LCoS SLMs \cite{Jesacher2014ColourRange}. The authors calibrate an SLM lookup-table consisting of phase-value triplets (for RGB) as a function of digital SLM input, and they note that SLMs with extended phase range (up to $10\pi$) can create substantial diversity in the calibrated phase triplets. After pre-optimizing a phase pattern for each color separately, the lookup-table is used on a per-pixel basis to find the digital input that best matches the desired phase for all colors. In our approach, we also use an extended SLM phase range for the same reason, but rather than using a two-step process, we directly optimize the output hologram. This flexible framework also allows us to incorporate a perceptual loss function to further improve perceived image quality.

\paragraph{Algorithms for Hologram Generation}
Our work builds on a body of literature applying iterative optimization algorithms to holographic displays. Perhaps most popular is the Gerchberg-Saxton (GS) method \cite{gerchberg1972practical}, which is effective and easy to implement, but does not have an explicitly defined loss function, making it challenging to adapt to specific applications. \citet{zhang20173d} and \citet{Chakravarthula2019WirtingerDisplays} were the first to explicitly formulate the hologram generation problem in an optimization framework. This framework has been very powerful, enabling custom loss functions \cite{Choi2022Time-multiplexedModulators} and flexible adaptation to new optical configurations \cite{choi2021optimizing, Gopakumar2021UnfilteredDisplays}. In particular, perceptual loss functions can improve the perceived image by taking aspects of human vision into account, such as human visual acuity \cite{kuo2020high}, foveated vision \cite{walton2022metameric}, and sensitivity to spatial frequencies during accommodation \cite{kim2022accommodative}. Like these prior works, we use an optimization-based framework which we adapt to account for the wavelength dependence of the SLM; this also enables our  new perceptual loss function for color, which is based on visual acuity difference between chrominance and luminance channels.

\paragraph{Camera-Calibration of Holographic Displays}
Mismatch between the computational model and physical system creates artifacts in experimental holograms. Recently, several papers have addressed this issue using measurements from a camera in the system for calibration.  %\citet{Peng2020NeuralTraining} proposed using feedback from the camera to update the SLM pattern for a particular image, and 
%although a single image can be improved, it does not extend to new content.
%a more flexible 
These approaches use pairs of SLM patterns and camera captures to estimate the learnable parameters in a model, which is then used for offline hologram generation. Learnable parameters can be physically-based \cite{Peng2020NeuralTraining, kavakli2022learned, Chakrabarti2016LearningBack-propagation}, black box CNNs \cite{Choi2021NeuralDisplays}, or a combination of both \cite{Choi2022Time-multiplexedModulators}. The choice of learnable parameters effects the ability of the model to match the physical system; we introduce a new parameter for modeling SLM cross talk and tailor the CNN architecture for higher diffraction orders from the SLM.  
% A special case of camera-calibration is what we'll refer to as ``active CiTL'' \cite{Peng2020NeuralTraining}. This is an online calibration in which a specific image is displayed, and feedback from the camera is used to update the SLM pattern for that particular image. Although active CiTL is only valid for a single image, it provides valuable proof-of-existence of experimental image quality. 

\section{Simultaneous Color Holography}
\label{ProblemStatement}

A holographic image is created by a spatially coherent illumination source incident on an SLM. The SLM imparts a phase delay on the electric field; after light propagates some distance, the intensity of the electric field forms an image. Our goal in this work is to compute a single SLM pattern that simultaneously creates an RGB hologram. For instance, when the SLM is illuminated with a red source, the SLM forms a hologram of the red channel of an image; with a green source the same SLM pattern forms the green channel; and with the blue source it creates the blue channel. 

We propose a flexible optimization-based framework (Fig.~\ref{fig:ForwardModel}) for generating simultaneous color holograms. We start with a generic model for estimating the hologram from the digital SLM pattern, $s$, as a function of illumination wavelength, $\lambda$:
% We can estimate the hologram from the digital SLM pattern, $s$, as a function of illumination wavelength, $\lambda$, as follows:
\begin{align}
    g_{\lambda} &= e^{i\phi_{\lambda}\left(s\right)} \label{eq:SLM_to_field} \\
    I_{z, \lambda} &= \left| f_{\text{prop}} \left( g_{\lambda}, z, \lambda \right) \right|^2. \label{eq:field_to_intensity}
\end{align}
Here, $\phi_{\lambda}$ is a wavelength-dependent function that converts the 8 bit digital SLM pattern to a phase delay, $g_{\lambda}$ is the electric field coming off the SLM, $f_{\text{prop}}$ represents propagation of the electric field, and $I_{z, \lambda}$ is the intensity a distance $z$ from the SLM.

To calculate the SLM pattern, $s$, we can solve the following optimization problem
\begin{align} \label{eq:loss_func_rgb}
    \argmin_{s} \sum_z \mathcal{L}\left(\hat{I}_{z, \lambda_r}, I_{z, \lambda_r}\right) + \mathcal{L}\left(\hat{I}_{z, \lambda_g}, I_{z, \lambda_g}\right) + \mathcal{L}\left(\hat{I}_{z, \lambda_b}, I_{z, \lambda_b}\right),
\end{align}
where $\hat{I}$ is the target image, $\mathcal{L}$ is a pixel-wise loss function such as mean-square error, and $\lambda_r, \lambda_g, \lambda_b$ are the wavelengths corresponding to red, green, and blue respectively. Since the model is differentiable, we solve Eq. \ref{eq:loss_func_rgb} with gradient descent.

% \begin{align}
%     \argmin_{s} \sum_{\lambda \in [\lambda_r, \lambda_g, \lambda_b]} \hspace{-5mm} \mathcal{L}\left(\hat{I}_{z, \lambda}, I_{z, \lambda}\right) 
% \end{align}

\begin{figure}
    \centering
    \includegraphics[clip, trim = 0in 7.7in 4.25in 0in ,width=.45\textwidth]{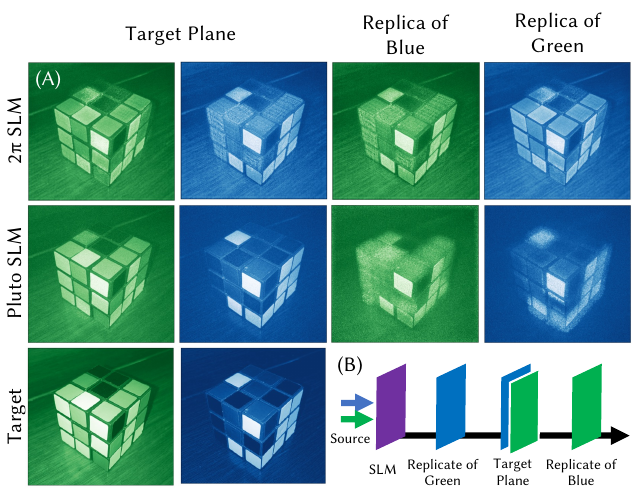}
    \caption{
    \textbf{Extended phase range reduces depth replicas in simulation.}
    (A) Using an SLM with a uniform $2\pi$ phase range across all channels leads to strong depth replicas (top row), which reduce image quality at the target plane compared to the target (bottom row) and add in-fcous content at depths that should be defocused. By using the extended phase Holoeye Pluto-2.1-Vis-016 SLM (with  Red: $2.4\pi$, Green: $5.9\pi$, Blue: $7.4\pi$ phase ranges), depth replicas are significantly reduced (middle row), improving the quality of target plane holograms and creating defocused content at other depths. (B) The illumination schematic illustrates the positions of the replicate planes and target plane. See Supplement for the three-color version of this figure. Rubik's cube source image by Iwan Gabovitch (CC BY 2.0).}
    \label{fig:replicates}
\end{figure}

\subsection{Color-Depth Ambiguity} \label{sec:color-depth-ambiguity}
A common model for propagating electric fields is Fresnel propagation\footnote{Fresnel propagation is the paraxial approximation to the popular angular spectrum method (ASM). Since most commercials SLMs have pixel pitch greater than \SI{3}{\micro\metre}, resulting in a maximum diffraction angle under $5^\circ$ (well within the small angle approximation), Fresnel and ASM are almost identical for holography.} \cite{goodmap2005fourier}, which can be written in Fourier space as
\begin{align}
    f_{\text{fresnel}}(g, z, \lambda) &= \mathcal{F}^{-1} \left\{ \mathcal{F}\{g\} \cdot H(z, \lambda)\right\} \label{eq:fresnel_convolution}\\
    H(z, \lambda) &= \exp \left(i \pi \lambda z \left(f_x^2 + f_y^2\right) \right) \label{eq:fresnel_kernel}% Frensel propagation kernel
\end{align}
where $\mathcal{F}$ is a 2D Fourier transform, $H$ is the Fresnel propagation kernel, and $f_x$, $f_y$ are the spatial frequency coordinates. In Eq.~\ref{eq:fresnel_kernel}, note that $\lambda$ and $z$ appear together, creating an ambiguity between wavelength and propagation distance.

To see how this ambiguity affects color holograms, consider the case where $\phi_{\lambda}$ in Eq. \ref{eq:SLM_to_field} is independent of wavelength ($\phi_{\lambda} = \phi$). For example, this would be the case if the SLM had a linear phase range from 0 to $2\pi$ at every wavelength.
%
%Although this is unrealistic for most off-the-shelf SLMs, such a device could be possible using geometric phase.
Although this is unrealistic for most off-the-shelf SLMs, it is a useful thought experiment.
Note that if $\phi$ is wavelength-independent, then so is the electric field off the SLM ($g_\lambda = g$). In this scenario, assuming $f_\text{prop} = f_\text{frensel}$, the Frensel kernel is the only part of the model affected by wavelength.

Now assume that the SLM forms an image at distance $z_0$ under red illumination. From the ambiguity in the Frensel kernel, we have the following equivalence:
\begin{align}
    H(z_0, \lambda_r) = H\left(\tfrac{\lambda_g}{\lambda_r} z_0, \lambda_g\right) = H\left(\tfrac{\lambda_b}{\lambda_r} z_0, \lambda_b\right).
\end{align}
This means the \textit{same} image formed in red at $z_0$ will also appear at $z = z_0 \lambda_g/\lambda_r$ when the SLM is illuminated with green and at $z = z_0 \lambda_b / \lambda_r$ when the SLM is illuminated with blue. We refer to these additional copies as ``depth replicas,'' and this phenomena is depicted in Fig. ~\ref{fig:replicates}.
Note that depth replicas do not appear in sequential color holography since the SLM pattern optimized for red is never illuminated with the other wavelengths.

If we only care about the hologram at the target plane $z_0$, then the depth replicas are not an issue. In fact, we can take advantage of the situation for hologram generation:
The SLM pattern for an RGB hologram at $z_0$ is equivalent to the pattern that generates a three-plane red hologram where the RGB channels of the target are each at a different depth ($z0$,  $z_0 \lambda_r/\lambda_g$, and  $z_0\lambda_r/\lambda_b$ for RGB respectively).
This is the basis of the depth division multiplexing approach of \citet{Makowski2008ColorfulHologram, Makowski2010ColorHolograms}, where the authors optimize for this three-plane hologram in red, then illuminate in RGB. Although this makes the assumption that $\phi$ does not depend on $\lambda$, this connection between simultaneous color and multi-plane holography suggests simultaneous color should be possible for a single plane, since multi-plane holography has been successfully demonstrated in prior work.

However, the ultimate goal of holography is to create 3D imagery, and the depth replicas could prevent us from placing content arbitrarily over the 3D volume.
In addition, in-focus images can appear at depths that should be out-of-focus, which may prevent the hologram from successfully driving accommodation \cite{kim2022accommodative}.
We propose taking advantage of SLMs with extended phase range to mitigate the effects of depth replicas.

\subsection{SLM Extended Phase Range} \label{sec:extended-phase}
In general, the phase $\phi_\lambda$ of the light depends on its wavelength, which was not considered in Sec. \ref{sec:color-depth-ambiguity}.
%In general, $\phi_\lambda$ depends on wavelength, unlike what we assumed in Sec. \ref{sec:color-depth-ambiguity}.
%
Perhaps the most popular SLM technology today is LCoS, in which rotation of birefringent LC molecules causes a change in refractive index.
The phase of light traveling through the LC layer is delayed by
\begin{align}
    \phi_\lambda = \frac{2 \pi d}{\lambda}n(s, \lambda), \label{eq:phase-LC}
\end{align}
where $d$ is the thickness of the LC layer, and its refractive index, $n$, is controlled with the digital input $s$.
$n$ also depends on $\lambda$ due to dispersion \cite{Jesacher2014ColourRange}.
% , which is particularly prominent in LC at blue wavelengths \cite{Jesacher2014ColourRange}.

The wavelength dependence of $\phi_\lambda$ presents an opportunity to reduce or remove the depth replicas.
Even if the propagation kernel $H$ is the same for several $(\lambda, z)$ pairs, if the phase, and therefore the electric field off the SLM, changes with $\lambda$, then the output image intensity at the replica plane will also be different.
As the wavelength-dependence of $\phi_\lambda$ increases, the replicas are diminished.

We can quantify the degree of dependence on $\lambda$ by looking at the derivative $d\phi / {d\lambda}$ which informs us that larger $n$ will give $\lambda$ more influence on the SLM phase.  
However, the final image intensity depends only on relative phase, not absolute phase;
therefore, for the output image to have a stronger dependence on $\lambda$, we desire larger $\Delta n = n_{\text{max}} - n_{\text{min}}$.
 In addition, $d\phi / {d\lambda}$ increases with $-dn / {d\lambda}$, suggesting that more dispersion is helpful for simultaneous color. Although $d\phi / {d\lambda}$ also depends on the absolute value of $\lambda$, we have minimal control over this parameter since there are limited wavelengths corresponding to RGB. In summary, this means we can reduce depth replicas in simultaneous color with larger phase range on the SLM and higher dispersion.

However, there is a trade-off: 
As the range of phase increases, the limitations of the bit depth of the SLM become more noticeable, leading to increased quantization errors.
We simulate the effect of quantization on hologram quality and find that PSNR and SSIM are almost constant for 6 bits and above \revised{(see Supplement)}.
This suggests that each $2\pi$ range should have at least 6 bits of granularity.
Therefore, we think that using a phase range of around $8\pi$ for an 8-bit SLM will be the best balance between replica reduction and maintaining accuracy for hologram generation.
Figure ~\ref{fig:replicates} simulates the effect of extended phase range on depth replica removal. While holograms were calculated on RGB images, only two color channels are shown for simplicity (see Supplement for full color version).
In the first row of Fig. ~\ref{fig:replicates}, we simulate an SLM
with no wavelength dependence to $\phi$ (i.e. 0 - $2\pi$ phase range for each color). Consequently, perfect copies appear at the replica planes.
In the second row, we simulate using the specifications from an extended phase range SLM (Holoeye Pluto-2.1-Vis-016), which has $2.4\pi$ range in red, $5.9\pi$ range in green, and $7.4\pi$ range in blue demonstrating that replicas are substantially diminished with an extended phase range.
By reducing the depth replicas, the amount of high frequency out-of-focus light at the sensor plane is reduced, leading to improved hologram quality.

\begin{figure}
    \centering
    \includegraphics[clip, trim = 0in 8.125in 4.27in 0in, width=.45\textwidth]{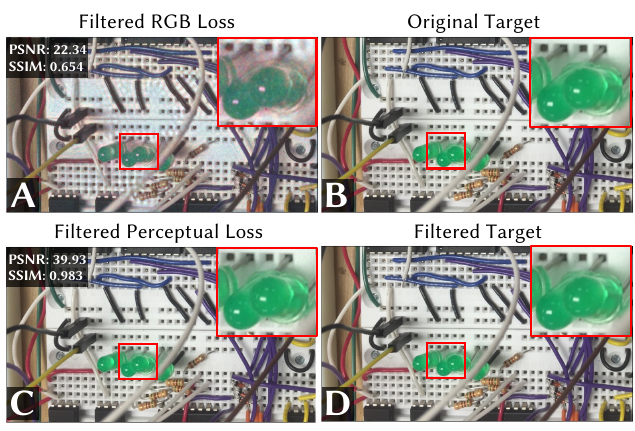}
    \caption{
    \textbf{Perceptual loss improves color fidelity and reduces noise in simulation.}
    The first column of this figure depicts simulated holograms optimized with an RGB loss function (A) and our perceptual loss function (C).  The same filters for the perceptual loss function then were applied to both of these simulated holograms as well as the target image.  Image metrics were calculated between the filtered holograms and the filtered target image (D). All image metrics are better for the perceptually optimized hologram (C). One should also note that the filtered target (D) and original target (B) are indistinguishable suggesting our perceptual loss function only removes information imperceptible by the human visual system.}
    \label{fig:perceptual_loss}
\end{figure}

\subsection{Perceptual Loss Function} \label{sec:perceptional-loss}

% taking out the statement about 3D being harder since we use a mask so there are the same number of target outputs with 3D as 2D.

Creating an RGB hologram with a single SLM pattern is an overdetermined problem as there are $3\times$ more output pixels than degrees of freedom of the SLM. As a result, it may not be possible to exactly match the full RGB image, which can result in color deviations and de-saturation. To address this, we take advantage of color perception in human vision. There's evidence that the human visual systems converts RGB images into a luminance channel (a grayscale image) and two chrominance channels, which contain information about the color~\cite{wandell1995foundations}. The visual system is only sensitive to high resolution features in the luminance channel, so the chrominance channels can be lower resolution with minimal impact on the perceived image \cite{wandell1995foundations}. This observation is used in JPEG compression \cite{pennebaker1992jpeg} and subpixel rendering \cite{platt2000optimal}, but to our knowledge, it has never been applied to holographic displays. By allowing the unperceived high frequency chrominance and extremely high frequency luminance features to be unconstrained, we can better use the the degrees of freedom on the SLM to faithfully represent the rest of the image.

Our flexible optimization framework allows us to easily change the RGB loss function in Eq. \ref{eq:loss_func_rgb} to a perceptual loss. For each depth, we transform the RGB intensities of both $\hat{I}$ (the target image) and $I$ (the simulated hologram) into opponent color space as follows: 
\begin{equation}
\begin{split}
    O_{1} &= 0.299 \cdot I_{\lambda_r} + 0.587 \cdot I_{\lambda_g} + 0.114 \cdot I_{\lambda_b} \\
    O_{2} &= I_{\lambda_r} - I_{\lambda_g}\\
    O_{3} &= I_{\lambda_b} - I_{\lambda_r} - I_{\lambda_g}
\end{split}
\end{equation}
where $O_1$ is the luminance channel, and $O_2$, $O_3$ are the red-green and blue-yellow chrominance channels, respectively. We can then update Eq. \ref{eq:loss_func_rgb} to
\begin{equation} \label{eq:loss_func_opponent}
\begin{split}
    \argmin_{s} \sum_z \Big[ &\mathcal{L}\left(\hat{O}_{1} * k_1, {O}_{1} * k_1 \right) + 
    \mathcal{L}\left(\hat{O}_{2} * k_2, {O}_{2} * k_2 \right) + \\
    &\mathcal{L}\left(\hat{O}_{3} * k_3, {O}_{3} * k_3 \right) \Big],
\end{split}
\end{equation}
where $*$ represents a 2D convolution with a low pass filter ($k_1 \hdots k_3$) for each channel in opponent color space . $\hat{O}_i$ and $O_i$ are the $i$-th channel in opponent color space of $\hat{I}$ and $I$, respectively. In order to mimic the contrast sensitivity functions of the human visual system, we implement filters in the Fourier domain by applying a low-pass filter of 45\% of the width of Fourier space to the chrominance channels ($O_2$, $O_3$) and a filter of 75\% of the width of Fourier space to the luminance channel ($O_1$).
In a system with a $\SI{36.6}{\milli\metre}$ focal length eye piece, these cutoffs  correspond to 30 cycles/deg and 18 cycles/deg in luminance and chrominance respectively, approximately matched to human vision~\cite{mullen1985contrast}.
%These filter widths were heuristically determined.

By de-prioritizing high frequencies in chrominance and extremely high frequencies in luminance, the optimizer is able to better match the low frequency color. This low frequency color is what is perceivable by the human visual system. Figure \ref{fig:perceptual_loss} highlights the improvement provided by our perceptual loss function, comparing perceptually filtered versions of simulated holograms generated using an RGB loss function (Fig \ref{fig:perceptual_loss}A) and our perceptual loss function (Fig \ref{fig:perceptual_loss}B). The original unfiltered target image (Fig \ref{fig:perceptual_loss}C) and the perceptually filtered target image (Fig \ref{fig:perceptual_loss}D) are nearly indistinguishable, indicating that our perceptual filter choices align well with the human visual system. The PSNR and SSIM values are higher for the perceptually optimized hologram (Fig. \ref{fig:perceptual_loss}C), which is visually less noisy with better color fidelity. This suggests that the loss function has effectively shifted most of the error into imperceptible regions of the opponent color space.  We see an average PSNR increase of 6.4 dB and average increase of 0.266 in SSIM across a test set of 294 images.

\subsection{Simulation Comparisons}

\begin{figure*}
    \centering
    \includegraphics[clip, trim = 0in 8.35in 0in 0in, width=.99\textwidth]{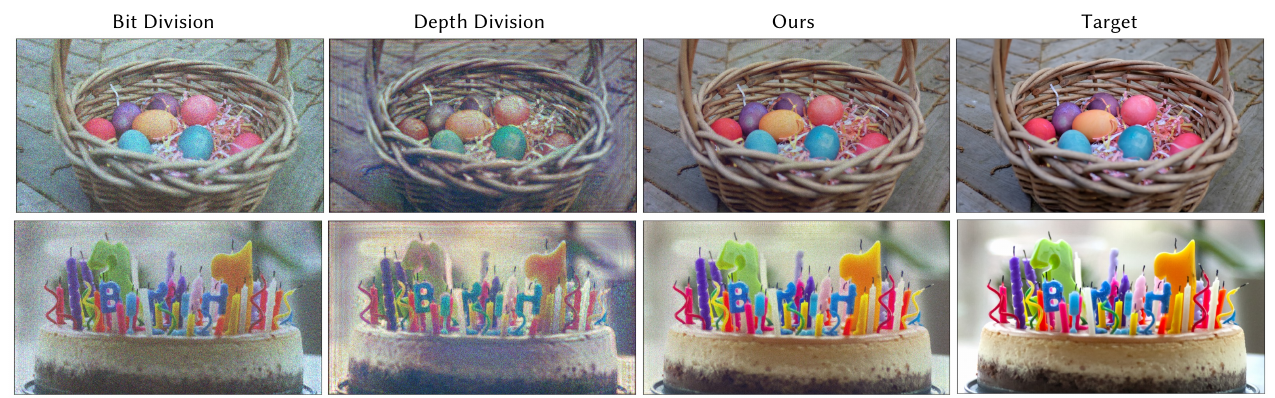}
    \caption{
    \textbf{Comparison of bit division, depth division and our method of simultaneous color holography in simulation.}
    Bit division (Col. 1) is noisier than our method (Col. 3) but achieves comparable color fidelity, although more washed out.  The depth division method (Col. 2) is also noisier than our method and has inferior color fidelity.  Our method matches the target  image (Col. 4) well.  Our method uses our perceptual loss function and a high order angular spectrum propagation model with no learned components. Further implementation details for each method are available in the supplement.
    }
    \label{fig:Simulation}
\end{figure*}

We compare the performance of our method to the depth and bit division approaches~\cite{Makowski2010ColorHolograms, Jesacher2014ColourRange}, which, like our method, use only a single SLM, make use of the full SLM space-time-bandwidth, and contain no bulky optics or filters (see Supplement for implementation details). The holograms simulated with depth and bit division, shown in Fig. ~\ref{fig:Simulation}, are much noisier and have lower color fidelity than our proposed method. Depth division has the worst color fidelity due to to the replica planes discussed in Sec. \ref{sec:color-depth-ambiguity} contributing defocused light at the target plane.  Our approach directly optimizes the simultaneous color hologram using our perceptual loss function, resulting in less noise and better color fidelity compared to these other indirect optimization approaches.

%Our method uses our perceptual loss function and the HOASM outlined by ~\citet{Gopakumar2021UnfilteredDisplays} to directly optimize the simultaneous color hologram, while comparison methods optimize indirectly.  This direct approach produces less noisy holograms with better color fidelity.

\section{Camera-Calibrated Model}\label{sec:learnable-model}

We've demonstrated that our algorithm can generate simultaneous color holograms in simulation. However, experimental holograms frequently do not match the quality of simulations due to mismatch between the physical system and the model used in optimization (Eqs. \ref{eq:SLM_to_field}, \ref{eq:field_to_intensity}). Therefore, to demonstrate simultaneous color experimentally, we need to calibrate the model to the experimental system. 

To do this, we design a model based on our understanding of the system's physics, but we include several learnable parameters representing unknown elements. To fit the parameters, we capture a dataset of SLM patterns and camera captures and use gradient descent to estimate the learnable parameters based on the dataset. Next we explain the model which is summarized in Fig. \ref{fig:ForwardModel}.

%\subsection{Learnable Parameters for Offline Calibration} 

\paragraph{Lookup Table} 
A key element in our optimization is $\phi_\lambda$ which converts the digital SLM input into the phase coming off the SLM. It's important this function accurately matches the behavior of the real SLM. Many commercial SLMs ship with a lookup-table (LUT) describing $\phi_\lambda$; however, this LUT is generally only calibrated at a few discrete wavelengths.  Consequently, we learn a LUT for each color channel's wavelength as part of the model. Based on a pre-calibration of the LUT using the approach of \citet{yang2015nonlinear}, we observe the LUT is close to linear; we therefore parameterize the LUT with a linear model to encourage physically realistic solutions.
% \begin{align}
%     \phi_lambda(s) = a_1 \cdot s + a_2
% \end{align}
% where $a_1, a_2$ are learned parameters that are different for each color channel. We initialize $a_1, a_2$ with the values approximated from our pre-calibration \gk{is this true?}.

\paragraph{SLM Crosstalk}
SLMs are usually modeled as having a constant phase over each pixel with sharp transitions at boundaries. However, in LCoS SLMs, 
elastic forces in the LC layer prevent sudden spatial variations, and the electric field that drives the pixels changes gradually over space. As a result, LCoS SLMs suffer from crosstalk, also called field fringing, in which the phase is blurred \cite{apter2004fringing, moser2019model, persson2012reducing}. We model crosstalk with a convolution on the SLM phase. Combined with our linear LUT described above, we can describe the phase off the SLM as
\begin{align} \label{eq:learned-model-01-LUT}
    \phi_\lambda(s) = k_{\text{xt}} * (a_1 \cdot s + a_2)
\end{align}
where $a_1, a_2$ are the learn parameters of the LUT, and $k_{\text{xt}}$ is a learned $5 \times 5$ convolution kernel representing crosstalk. Separate values of these parameters are learned for each color channel.

\paragraph{Propagation with Higher Diffraction Orders}
The discrete pixel structure of the SLM creates higher diffraction orders that are not modeled by ASM or Fresnel propagation. With the use of a 4$f$ system, a physical aperture at the Fourier plane of the SLM can be used to block higher orders. However, this adds significant size to the optical system, reducing the practicality for head-mounted displays. Therefore, we chose to avoid additional lenses after the SLM and instead account for higher orders in the propagation model.

We adapt the higher-order angular spectrum model (HOASM) of \citet{Gopakumar2021UnfilteredDisplays}. 
% Defining $\mathcal{F}\{g\} = G(f_x, f_y)$, where $g$ is the electric field off the SLM from Eq. \ref{eq:SLM_to_field}, we calculate the Fourier transform of the first diffraction orders as
% \begin{equation} \label{eq:learned-model-02-HOASM}
% \begin{split}
%     G_{\text{1st order}}(f_x, f_y) = &\left(- G(f_x, f_y) + \sum_{i,j = -1}^1 G \left(f_x + \tfrac{i}{p}, f_y + \tfrac{j}{p} \right)  \right)  \\ & \times \text{Sinc}(\pi f_x p) \cdot \text{Sinc}(\pi f_y p)
% \end{split}
% \end{equation}
% where $p$ is the pixel pitch of the SLM. 
The zero order diffraction, $G_0(f_x, f_y)$, and first order diffraction, $G_1$, patterns are propagated with ASM to the plane of interest independently. Then the propagated fields are stacked and passed into a U-net, which combines the zero and first orders and returns the image intensity:
\begin{align} 
f_{\text{ASM}}(G, z) &= \mathcal{F}^{-1}\left\{ G \cdot H_{\text{ASM}}(z)\right\} \label{eq:learned-model-03-propagation} \\ 
I_z &= \text{Unet}\left(f_\text{ASM}(G_0, z), \: f_\text{ASM}(G_1, z)\right), \label{eq:learned-model-04-unet}
\end{align}
where $H_{\text{ASM}}(z)$ is the ASM kernel. The U-Net architecture is detailed in the supplement; a separate U-net for each color is learned from the data. The U-Net helps to address any unmodeled aspects of the system that may affect the final hologram quality such as source polarization, SLM curvature, and beam profiles, and the U-net better models superposition of higher orders, allowing for more accurate compensation in SLM pattern optimization.  

\section{Implementation}

\begin{figure*}
    \centering
    \includegraphics[clip, trim = 0in 7.75in 0in 0in, width=.9\textwidth]{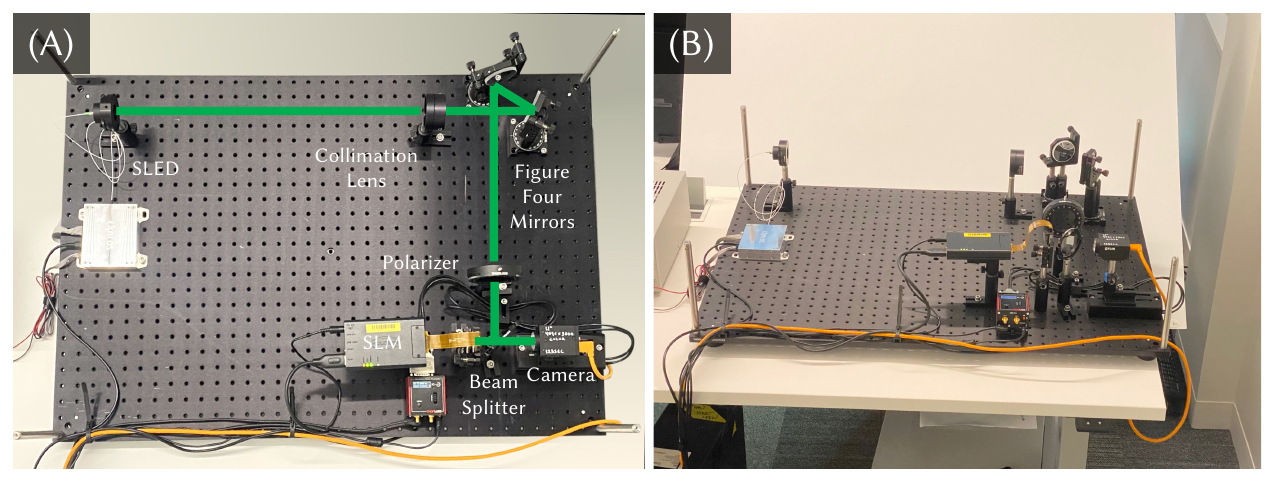}
    \caption{\textbf{Experimental setup} A top view of our system with labeled components and an approximate beam path drawn in green is depicted in (A).  A side-view of the system is provided by (B). \revised{Note that the hologram is formed directly on the bare camera sensor with no lens or eyepiece between. This configuration allows us to validate our method, but for a human-viewable system, an eyepiece must be added between the hologram plane and the user's eye.}
    }
    \label{fig:Setup}
    \end{figure*}

\paragraph{Experimental Setup}
Our system starts with a fiber-coupled RGB source ($\lambda_r = \SI{636}{\nano\metre}, \lambda_g = \SI{512}{\nano\metre}, \lambda_b = \SI{453}{\nano\metre})$, collimated with a $\SI{400}{\milli\metre}$ lens. The beam is aligned using two mirrors, passes through a linear polarizer and beamsplitter, reflects off the SLM (Holoeye-2.1-Vis-016), and passes through the beamsplitter a second time before directly hitting the color camera sensor with Bayer filter (FLIR GS3-U3-123S6C). As seen in Fig.~\ref{fig:Setup}, there's no 4$f$ system between the SLM and camera, which allows the setup to be compact, but requires modeling of higher diffraction orders. The camera sensor is on a linear motion stage, enabling a range of propagation distances from $z = \SI{80}{\milli\metre}$ to $z = \SI{130}{\milli\metre}$.

For our source, we use a superluminescent light emitting diode (SLED, Exalos EXC250011-00) rather than a laser due to its lower coherence, which has been demonstrated to reduce speckle in holographic displays \cite{Deng2017CoherenceDisplays}. \revised{Although previous work showed state-of-the-art image quality by modeling the larger bandwidth of the SLED as a summation of coherent sources \cite{Peng2021Speckle-freeCalibration}, we found the computational cost to be prohibitively high for our application due to GPU memory constraints. We achieved sufficient image quality while assuming a fully coherent model, potentially due to the U-net which is capable of simulating the additional blur we expect from a partially coherent source.}

\revised{Our experimental system directly forms the hologram on a bare sensor, but for a human-viewable system, an eyepiece is necessary between the image plane and the user's eye. See Supplement for details on how the eyepiece effects the depth replicas.}

\paragraph{Calibration Procedure}
We learn parameters in our model (Eqs. \ref{eq:learned-model-01-LUT} - \ref{eq:learned-model-04-unet}) using a dataset captured on the experimental system. We pre-calculate 882 SLM patterns from a personally collected dataset of images using the ASM propagation model. Each SLM pattern is captured in $\SI{10}{\milli\metre}$ increments from $z = \SI{90}{\milli\metre}$ to $\SI{120}{\milli\metre}$%, resulting in a total of 6174 paired entries. 
The camera data is debayered and an affine transform is applied to align the image with the SLM (see Supplement for details). Model fitting is implemented in PyTorch using an L1 loss function between the model output and camera capture. To account for the camera color balance, we additionally learn a $3 \times 3$ color calibration matrix. % from the RGB simulated intensities to the captured color image.
We train until convergence, which is typically reached in 2-3 days on Nvidia A6000 GPU.

\paragraph{Hologram Generation}
After training, we can generate holograms by solving Eq. \ref{eq:loss_func_opponent} using the trained model for $I_{z,\lambda}$, implemented with PyTorch's native autodifferentiation. The SLM pattern, $s$, is constrained to the range where the LUT is valid (for example, 0 - 255); the values outside that range are wrapped after every optimization step.  On the Nvidia A6000 GPU, it takes about two minutes to optimize a 2D hologram.  Computation time for the optimization of a 3D hologram scales proportionally to the number of depth planes.

\section{Experimental Results}

\paragraph{2-Dimensional Holograms}
We validate our simulation results by capturing holograms in experiment. For simultaneous color, the SLM patterns were optimized for a propagation distance of \SI{120}{\milli\metre} using our perceptual loss function described in Section \ref{sec:perceptional-loss}. A white border was added to each target image to improve the color fidelity by encouraging a proper white balance. After each hologram is captured, debayering is performed and a homography is applied to map from camera space to SLM space.

\begin{figure*}
    \centering
    \includegraphics[clip, trim = 0in 7.1in 0in 0in, width=0.95\textwidth]{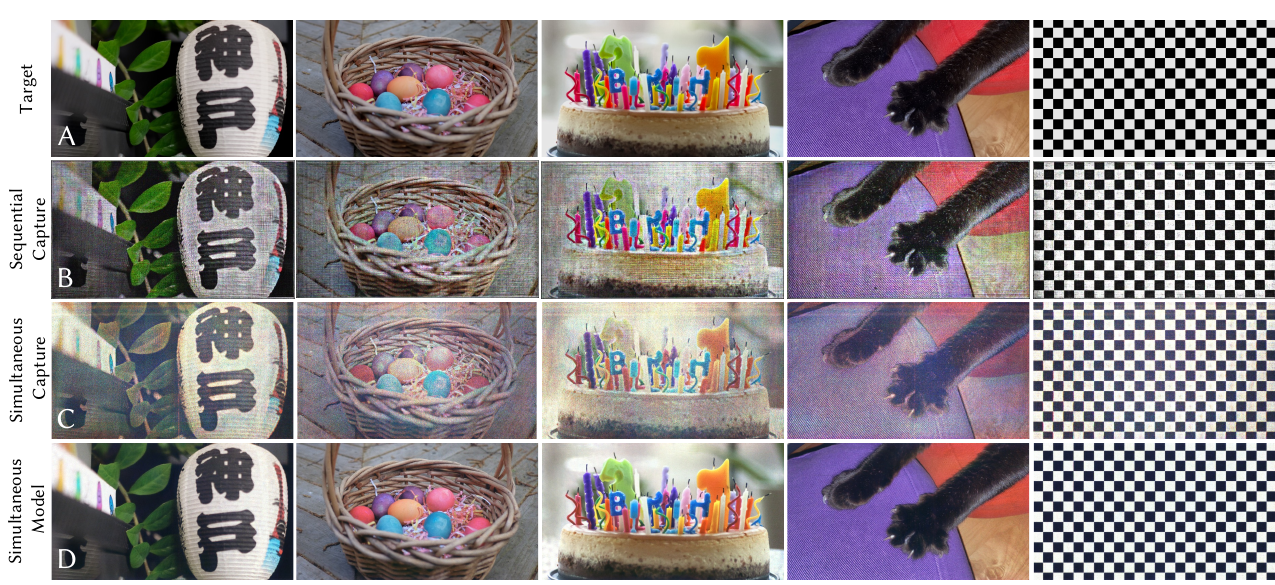}
    \caption{\textbf{Experimentally captured 2D holograms.}
    For each target image (A), we show (B) the experimental capture with sequential pseudo-color, (C) our experimental capture with full simultaneous color, and (D) the simulated model output for simultaneous color. Recall that our simultaneous color results (C) use 3$\times$ fewer degrees of freedom than the sequential capture (B). Although some color fidelity is lost in experiment (C), the simulated model output (D) shows good color quality, demonstrating that accurate color is possible with our method and improvements to the calibration.
    % and the which was captured in pseudo-color using the red light source only. 
    % %
    % This figure shows experimentally captured holograms at a depth of \SI{120}{\milli\metre}. Row one: target images. Row two: experimentally captured sequential color holograms. Row three: experimentally captured simultaneous color holograms. Row four: simulation output of the optimized simultaneous color SLM pattern. While most captured simultaneous color holograms have good color fidelity, our method is least effective on highly saturated images with low texture, such as the cat paws in column 4, representing a limitation of our method (see Sec. ~\ref{sec:limits}).
    }
    \label{fig:Stills}
    \end{figure*}

Figure~\ref{fig:Stills} compares the simultaneous color capture using a single frame (B) to sequential color using 3 frames (C). Unlike the simultaneous color version, which was captured in one shot with RGB illumination, the sequential color was captured with only the red light source (due to a failure of the green channel in the SLED), and the correct color was assigned in software. Although the sequential captures are higher contrast than our simultaneous results, we'd like to emphasize that our approach uses 3$\times$ fewer degrees of freedom and can still produce full color images. In addition, the simulation output from our model (D) shows color fidelity on par with the sequential capture; the difference between the simulation output and experimental capture can be attributed to model mismatch. This suggests improvements to the calibration pipeline could enable experimental results with the  quality of the simultaneous model.

\begin{figure*}
    \centering
    \includegraphics[clip, trim = 0in 6.42in 0in 0in, width=0.95\textwidth]{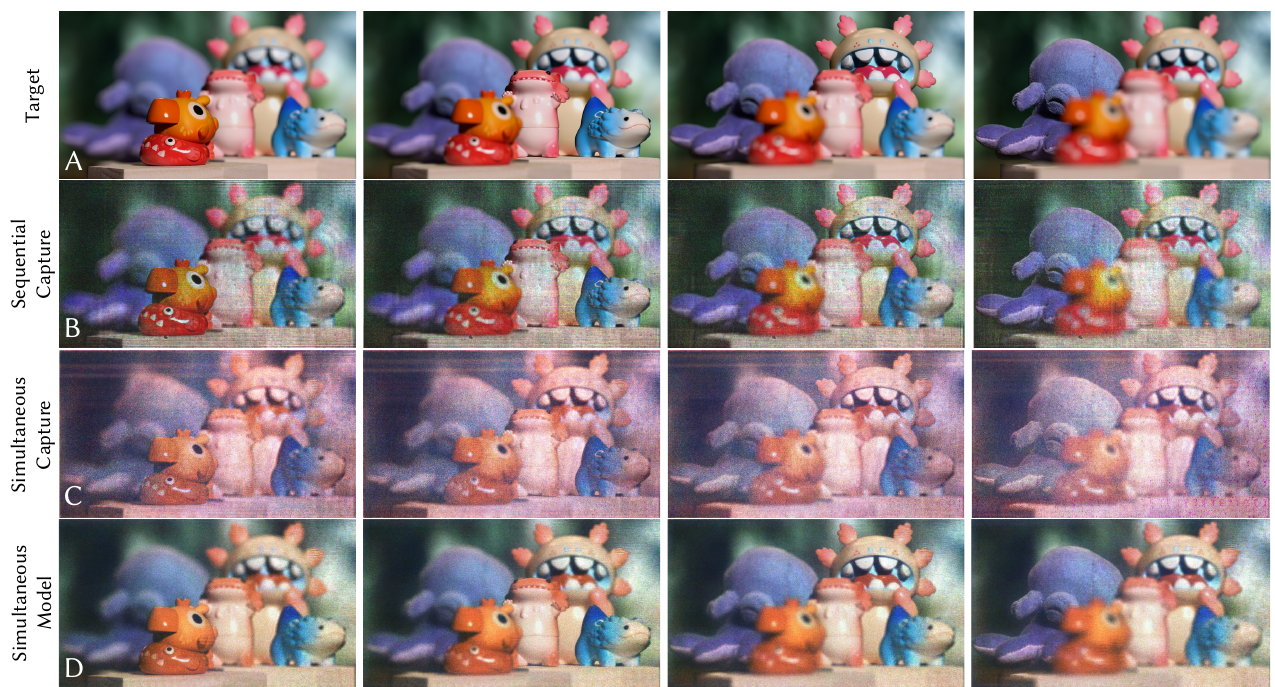}
    \caption{\textbf{Experimentally captured focal stack.} This figure displays a focal stack, with the target shown in (A), captured from $\SI{90}{\milli\metre}$ to $\SI{120}{\milli\metre}$ in $\SI{10}{\milli\metre}$ increments. We compare (B) the sequential pseudo-color experimental capture with (C) the experimental capture of the simultaneous full color hologram and (D) the simulated model output for simultaneous color. Although model mismatch creates some deviations between the simultaneous capture (C) and the target (A), the simulated model (D) is representative of the color fidelity we expect from our method with improvements to the system calibration.
    }
    \label{fig:Stack}
    \end{figure*}

\paragraph{3-Dimensional Holograms}
A major appeal of holography is the ability to solve the vergence-accommodation conflict, so we also validate our method for 3D scenes.  A 4-plane focal stack was rendered with 0.5 pixels blur radius per millimeter depth.  Holograms were captured at distance from \SI{90}{\milli\metre} to \SI{120}{\milli\metre} in \SI{10}{\milli\metre} increments. The results are displayed in Fig. \ref{fig:Stack}, and once again pseudo-color sequential images (B), which use $3\times$ the number of frames, are shown for comparison. \revised{Although model mismatch creates some color shift in the experimental captures (C), the simultaneous model output (D) shows what the results could look like with improved calibration. We note that 3D hologram generation is not as well-posed as 2D;} despite this, our results demonstrate the ability to form 3D color holograms with natural defocus blur from a single SLM frame.

%Color sequential images, which use $3\times$ the number of frames, are are included as a point of comparison and are captured in the same manner as the 2D case. 

\section{Discussion}

% \section{Limitations}
While our method improves hologram quality for simultaneous illumination and is compatible with VR/AR displays, it does have limitations.  First, our method is not equally effective for all images.  Natural images with high levels of texture work best, as they have similarly structured color channels and contain high frequency color information that is perceptually suppressible by our loss function. 
Images with large flat areas may exhibit noticeable artifacts due to the more difficult task of determining an SLM pattern that produces 3 largely unique holograms \revised{(see Supplement Fig. S4)}.  

SLMs with large phase range can be slower than their short phase range counterparts. Although our SLM has $7.4\pi$ phase range in blue, we show in the Supplement that we can achieve reasonable quality with only a $4\pi$ range, opening the possibility for simultaneous color with a wider variety of SLMs.

Calculating a single SLM pattern for a 2D image using an Nvidia A6000 takes minutes with our method, inhibiting real-time displays. Neural nets can generate SLM patterns in real-time while retaining quality, suggesting a potential future solution for simultaneous color holography~\cite{shi2021towards, eybposh2020high, yang2022diffraction}.

\label{sec:limits}

% Finally, many recent works have shown active camera-in-the-loop (CiTL) can dramatically increase hologram quality.  While this is true, using active CiTL for a user display is not feasible.  As the argument for simultaneous color is to make color holographic displays possible for VR and AR, we have chosen to not focus on active CiTL in this work.  Active CiTL does provide proof-of-existence of experimental image quality and our method is compatible with it.  Results using active CiTL can be found in the supplement. 
% \paragraph{Conclusion}
% In conclusion, this work presents a comprehensive framework for the generation of color holograms using simultaneous RGB illumination with a simple and compact optical setup. The framework utilizes a camera-calibrated and differentiable forward model, which reduces model mismatch and enables the use of custom loss functions. A perceptual loss function is employed to address the over-determined problem of simultaneous color holography, resulting in higher quality results. All findings were validated through experimental testing, bringing us closer to achieving the goal of creating holographic near-eye displays.

In summary, we developed a framework for high-quality color holograms using simultaneous RGB illumination in a compact setup, featuring a camera-calibrated, differentiable model and custom loss functions. %Our perceptual loss function effectively addresses the difficult challenges of simultaneous color holography, proven by successful 2D and 3D experiments, advancing us towards holographic near-eye displays.

\clearpage 

\begin{figure*}
    \centering
    \includegraphics[clip, trim = 0in 6.125in 0in 0in, width=\textwidth]{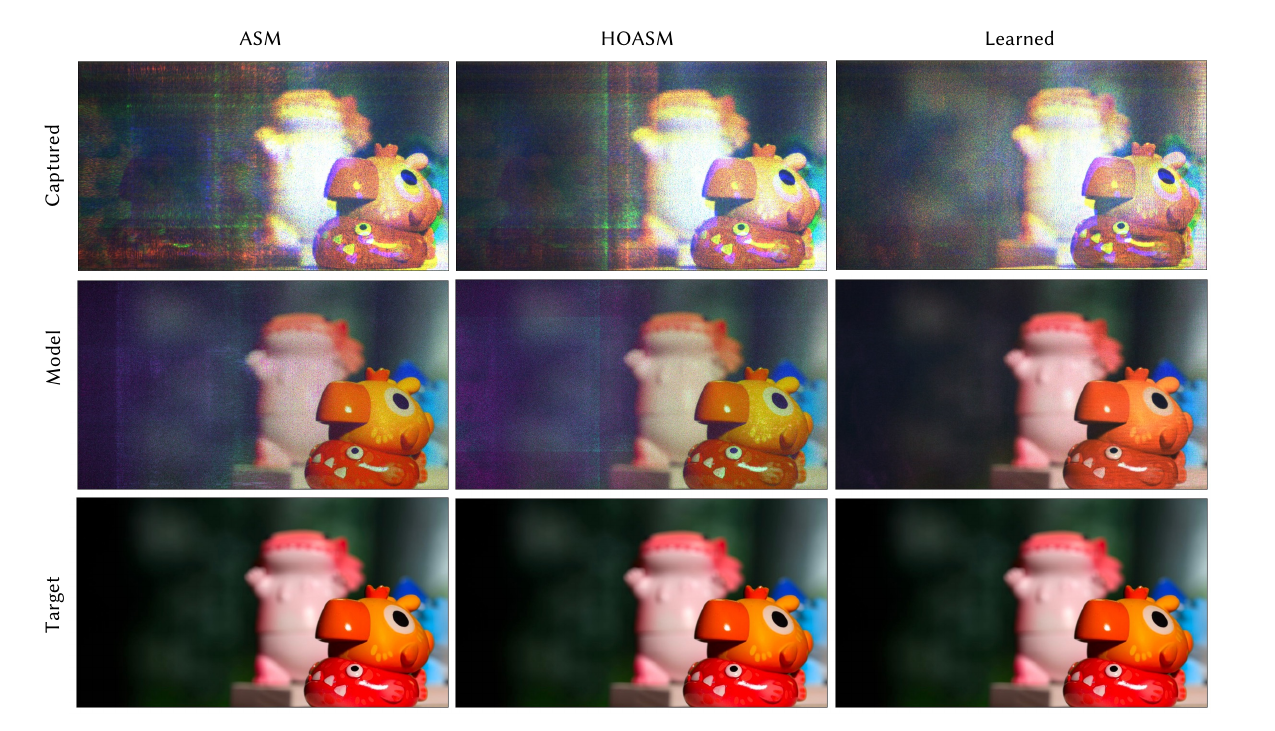}
    \caption{\textbf{Comparison of different propagation methods for suppressing higher diffraction orders.} The first column shows the results obtained using the traditional angular spectrum method (ASM) which doesn't model higher diffraction orders. The second column shows the results obtained using HOASM which reduces the visibility of higher orders but fails to completely suppress them. The third column shows the results obtained using our proposed learned propagation method that includes a U-net, which largely suppresses the higher diffraction orders and results in a hologram with the fewest artifacts, suggesting the learned propagation model best matches the physical propagation.
    }
    \label{fig:Ablation}
    \end{figure*}

\clearpage

%%%%%%%%%%%%%%%%%%%%%%%%%%%%%%%%%%%%%%%%%%%

%\clearpage

\bibliographystyle{ACM-Reference-Format}
\bibliography{references_local}

\clearpage

\newcommand{\beginsupplement}{%
    \setcounter{table}{0}
    \renewcommand{\thetable}{S\arabic{table}}%
    \setcounter{figure}{0}
    \renewcommand{\thefigure}{S\arabic{figure}}%
    \setcounter{section}{0}
    \renewcommand{\thesection}{S\arabic{section}}%
 }

\beginsupplement{}
\onecolumn
\textbf{\huge Supplementary Material -- Simultaneous Color Holography}

\section{Additional Implementation Details}

\paragraph{Spatial Light Modulator}
For all simulations, a spatial light modulator (SLM) with $1920 \times 1080$ pixels and a pixel size of $\SI{8}{\micro\metre} \times \SI{8}{\micro\metre}$ is used. The phase ranges of the red, green, and blue channels are $2.4\pi$, $5.9\pi$, and $7.4\pi$, respectively, unless otherwise noted. These values were experimentally calibrated for the Holoeye Pluto-2.1-Vis-016 SLM. The propagation distance of all simulated holograms is $\SI{100}{\milli\metre}$ unless otherwise noted.

\paragraph{Modified High Order Angular Spectrum Method (HOASM)}
We implement a modified version of the High Order Angular Spectrum Method (HOASM) as described by \citet{Gopakumar2021UnfilteredDisplays}. Instead of propagating the zero- and first-order together, we propagate them separately. The zero-order is propagated by performing the traditional angular spectrum method (ASM). To propagate the first-order, we pattern the zero-padded Fourier transform of the complex field to be propagated into a $3 \times 3$ grid. The center Fourier transform of the grid is then zeroed out. The Fourier representation of the first-order is then weighted with a sinc function and propagated to the sensor plane using ASM. The field is then down-sampled and cropped. The complex fields of the zero- and first-orders are then split into real and imaginary parts and stacked before being fed into a U-Net. The U-Net consists of 4 downsampling layers, the number of channels increases from 4 to 32 during the first downsampling layer and doubles in each of the next 3 downsampling layers until there are 256 channels. Four upsampling layers are then applied, producing a single-channel output representing the intensity of the propagated wavefront.

\paragraph{ Camera Space to SLM Space Homography}
To perform either offline or active camera-in-the-loop optimization, the captured wavefront and SLM must be in the same space. This requires a transform and downsampling of the captured image to place it in the same coordinate system as the SLM pattern used to generate it. We opt to use an affine transform to perform this mapping. The affine transform is calculated as follows: first, an SLM pattern is calculated that produces a grid of dots. The dots are then detected on the sensor, and their centers are estimated in camera space coordinates. The centers of the dots are known in SLM space since the target image containing the dots is in SLM space for optimization. Finally, Python's OpenCV package is used to produce the affine transform matrix that maps the captured dots to the SLM coordinate space. A unique homography is calculated for each depth location and color channel.

\paragraph{Source Power Optimization}
Correctly setting the power of each color channel of the SLED for a given hologram is an important step to achieving good color fidelity. To achieve this, we use an active camera-in-the-loop based approach to optimize the power of the color channels. First, the power of the source is set to an arbitrary value less than 100\% across all three color channels. A baseline reference image is captured, debayered, and mapped to the SLM space. Three learnable weighting parameters, one for each color channel, are initialized to unity and applied to the captured reference image. These weighting parameters serve as a proxy to optimizing the source power. An iterative process is then undertaken, where an image is captured on the camera, debayered, and mapped to the SLM space. The loss between this image and the target image is calculated and then backpropagated using the computational graph of the weighting parameters applied to the reference image. The initial source power is then multiplied by the updated weighting parameters, and a new image is captured, restarting the iterative loop. This is done until the color weighting parameters have converged, usually taking between 15-30 iterations. If the process fails to converge or the initial source power multiplied by the weighting function becomes greater than 100\%, the exposure time is increased, and the source power optimization is restarted.

Although we use camera feedback in this process, we note that the information needed for source power optimization is contained in the color balance of the image itself. We believe this step could be replaced with a precomputed source power that's dependent on the image content.

\paragraph{Simultaneous Color Focal Stack Optimization} 
In this approach, we optimize the SLM pattern for a focal stack using simultaneous color holography. First, we load a learned model for wave propagation for our unfiltered holography system. Then, the SLM pattern is initialized and target intensities defined for each plane in the multiplane hologram. Gradient descent is used to optimize the SLM pattern. For each plane, the field generated from the current SLM pattern is calculated and propagated to the current plane of interest using the learned propagation model. The loss is computed using our custom perceptual loss function and back-propagated to calculate the gradient of the loss with respect to the SLM pattern. The calculated gradients are then used to update the SLM pattern.  This process is performed iteratively for each depth plane and continued until the SLM pattern converges. Finally, we capture a simultaneous color focal stack by displaying the optimized SLM pattern, moving the camera to each plane in the multiplane hologram and capturing.  Pseudocode for this method is provided in Algorithm \ref{alg:3d_holography}.

\begin{algorithm}[H]
\caption{Simultaneous Color Focal Stack Optimization}
\label{alg:3d_holography}
\begin{algorithmic}[1]

\State \textbf{Initialize:} Load learned model for wave propagation. Initialize SLM pattern $P_{SLM}$ and target intensity $I_{target}$ for each depth plane in the focal stack.

\State \textbf{Optimize:} \While{SLM pattern $P_{SLM}$ not converged}

    \For{each depth plane, $d$, in the focal stack hologram}
        \State Clear previous gradients
        \State Generate field from current SLM pattern and propagate to depth plane $d$: $Field_{out} = Holo(P_{SLM}, depth = d)$
        \State Compute loss with custom perceptual loss function: $Loss = LossFunc(Field_{out}, I_{target}[d])$
        \State Backpropagate the gradient of $Loss$ with respect to $P_{SLM}$
        \State Update the SLM pattern $P_{SLM}$
    \EndFor

\EndWhile

\State \textbf{Capture:} Experimentally capture the simultaneous color focal stack
    \For{each depth plane, $d$, in the focal stack hologram}
        \State Move camera to depth plane $d$
        \State Capture hologram: $Hologram = CaptureImage(camera)$
        \State Add captured hologram to the focal stack: $FocalStack.append(Hologram)$
    \EndFor

\State \textbf{Output:} The optimized SLM pattern, $P_{SLM}$, and the captured simultaneoius color focal stack, $FocalStack$

\end{algorithmic}
\end{algorithm}
\clearpage

\revisedb{
\section{Effect of Eyepiece on Near-eye Holographic Displays}
Our experimental setup creates an image directly on the bare sensor, but for a human-viewable system, we would use an eyepiece between the image plane and the user's eye. For a near-eye display, the eyepiece is needed to make the image plane appear further away so the eye can focus on it. Together with the lens of the user's eye, the eyepiece creates an optical relay system that generates a copy of the image plane on the user's retina.}

\revisedb{
There are three major differences between this more realistic setup with an eyepiece and our experimental setup without an eyepiece: (1) The relay system generally does not have unit magnification, which causes the image plane and corresponding color depth replicas to appear at new axial locations. (2) The finite aperture of the eyepiece can block some of the wavefront, creating artifacts and vignetting at the edges of the field of view. (3) Optical aberrations in the eyepiece can reduce image quality, adding blur or undesirable speckle if not compensated for.}

\revisedb{
We now explore each of these differences in more detail.}

\subsection{Effect of the Eyepiece on Depth Replica Locations}

After being viewed through the eyepiece, the image plane of the holographic display appears at a new depth in the world. This is governed by the thin lens equation
\begin{equation}
    z_{\text{world}} = \frac{f z_{\text{slm}}}{f - z_{\text{slm}}},
    \label{eq:thin_lens}
\end{equation}
where $f$ is the focal length of the eyepiece, $z_{\text{slm}}$ is the actual distance from the eyepiece to the image plane, and $z_{\text{world}}$ is the apparent distance to the image when viewed through the eyepiece.

\revised{
The apparent positions of the replica planes will also be shifted in the world based on the eyepiece, and in general, replica planes are spread further apart. We illustrate this effect with a concrete example.
}

\revised{
Consider the case where we have an eyepiece of focal length $f = \SI{30}{\milli\meter}$ and the SLM is co-located with the eyepiece for a thin form factor. If we want the image plane to appear at 1 m in the world, we can use Eq.~\ref{eq:thin_lens} to calculate that the propagation distance from the SLM should be $z_0 = \SI{29.13}{\milli\metre}$. If the image is created in green (512 nm), the replicas in red (636 nm) and blue (453 nm) will appear at \SI{36.18}{\milli\metre} and \SI{25.77}{\milli\metre}, respectively, relative to the SLM. For both of these replica planes, we can apply Eq.~\ref{eq:thin_lens} to calculate the position of these planes in the world after being viewed through the eye piece.}

\revised{
For the red replica at \SI{36.18}{\milli\metre}, $z_{\text{world}}$ is negative, indicating that the apparent image is \textit{behind} the viewer (in other words, the illumination is converging, rather than diverging)---this is not representative of real, physical objects and our eyes generally cannot focus on this plane, meaning that this replica is no longer visible.}

\revised{
For the replica in blue at \SI{25.77}{\milli\metre} relative to the SLM, we calculate that this corresponds to a depth 18.28~cm once viewed through the eyepiece. Recall that the original (non-replica) plane is a 1 m from the eye piece, demonstrating that the eyepiece causes the replicas to be significantly more spread out. In fact, the eyes of many adults cannot accommodate at such a close distance, meaning that such individuals would not be able to focus on the replica plane.
}

\revised{
In general, as the eyepiece focal length gets shorter, the small distances between the replicas are exaggerated. This makes them less visible since the replicas appear in regions where it's more difficult for the viewer to accommodate. 
}

\clearpage
\begin{figure*}
    \centering
    \includegraphics[width=\textwidth]{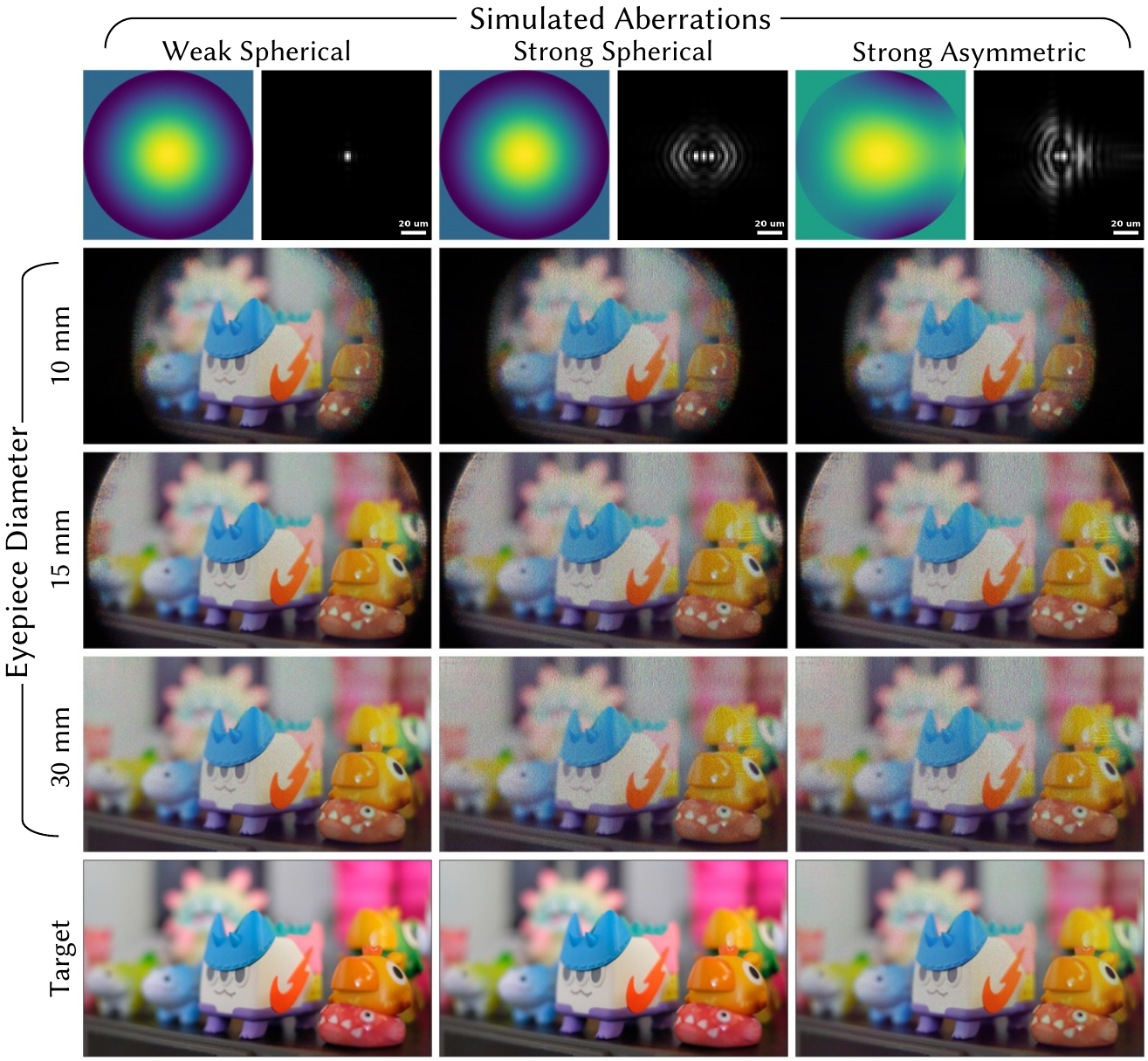}
    \caption{\revisedb{\textbf{Effect of eyepiece on image quality (simulation).} Each column shows different simulated eyepiece aberrations, with the Zernike phase error and PSF shown at the top. Each row shows a different eyepiece diameter. With small aberrations where the PSF is contained to a single SLM pixel (left column), there is minimal impact on image quality. Stronger aberrations, both symmetric (center column) and asymmetric (right column) add additional speckle to the image. This could be compensated for computationally if the aberrations are known. When the eyepiece diameter is smaller than the SLM size (10 mm, top) only the center of the image is visible and there is additional speckle at the edges. When the eyepiece is approximately the same size as the SLM (15 mm, center), there is vignetting in the corners and reduced quality at the edges. When the eyepiece is sufficiently large (30 mm, bottom), there are no artifacts from the eyepiece aperture size.}} 
    \label{fig:eyepiece}
\end{figure*}
\clearpage

\revisedb{
\subsection{Effect of the Eyepiece on Image Quality}
}

\revisedb{
Since the eyepiece creates a relay system between the image plane and the user's retina, with an ideal eyepiece, the image seen by the user is a magnified version of the image directly in front of the SLM. However, the finite size of the eyepiece aperture can cause artifacts at the edges of the field of view, and aberrations in the eyepiece can further reduce image quality.}

\revisedb{
Figure~\ref{fig:eyepiece} simulates the effect of these non-idealities on image quality. We assume an $f=\SI{30}{\milli\metre}$ focal length eyepiece and an SLM with \SI{8}{\micro\metre} pixels and $1920 \times 1080$ resolution, which matches our experimental results. We simulate an eyepiece diameter ranging from \SI{10}{\milli\metre} to \SI{30}{\milli\metre}. Note that the SLM is about \SI{16}{\milli\metre} across. When the eyepiece diameter is larger than the SLM size (30 mm, bottom), there are no visible edge effects in the image. When the eyepiece diameter is close to the SLM size (15 mm, middle), some of the image is lost in the corners and there's additional speckle around the edges. When the eyepiece has a significantly smaller diameter (10 mm, top), only a fraction of the image is viewable due to the finite size of the eyepiece. This suggests that the eyepiece diameter must by larger than the size of the image plane in order to cover the full field of view.}

\revisedb{
We also simulate the effect of aberrations on image quality. The left column show a simulated lens with a small amount of spherical aberration, which is representative of a well-corrected eyepiece. Since the point spread function (PSF) of this simulated lens is smaller than \SI{8}{\micro\metre}, the SLM pixel size, we see good image quality with minimal speckle. Once the spot size starts to exceed the pixel size of the SLM, additional speckle becomes visible in the image. The image quality is qualitatively similar for symmetric aberrations like spherical (center column) and asymmetric aberrations (right column). We note that holographic displays can compensate for aberrations \cite{maimone2017holographic} if the aberrations are known, so even poorly corrected eyepieces can yield high quality images. However, here we simulate the case where aberrations are not corrected for computationally.}

\revisedb{
In conclusion, without further computational correction, an eyepiece will not reduce image quality if the PSF is smaller than the SLM pixel size and the eyepiece diameter is larger than the total SLM size.}
\clearpage

\section{Active camera-in-the-Loop (CiTL)}
Active CiTL \cite{Peng2020NeuralTraining} is a special case of camera-calibrated models in which an image is displayed on the SLM, and camera captures are used to improve that particular image using the difference between the experimental capture and target image. While active CiTL is incompatible for real time displays, it does provide a useful proof of achievable hologram quality. Consequently, we implemented active CiTL for our system as follows.

\begin{figure*}
    \centering
    \includegraphics[clip, trim = 0in 3in 0in 0in, width=0.9\textwidth]{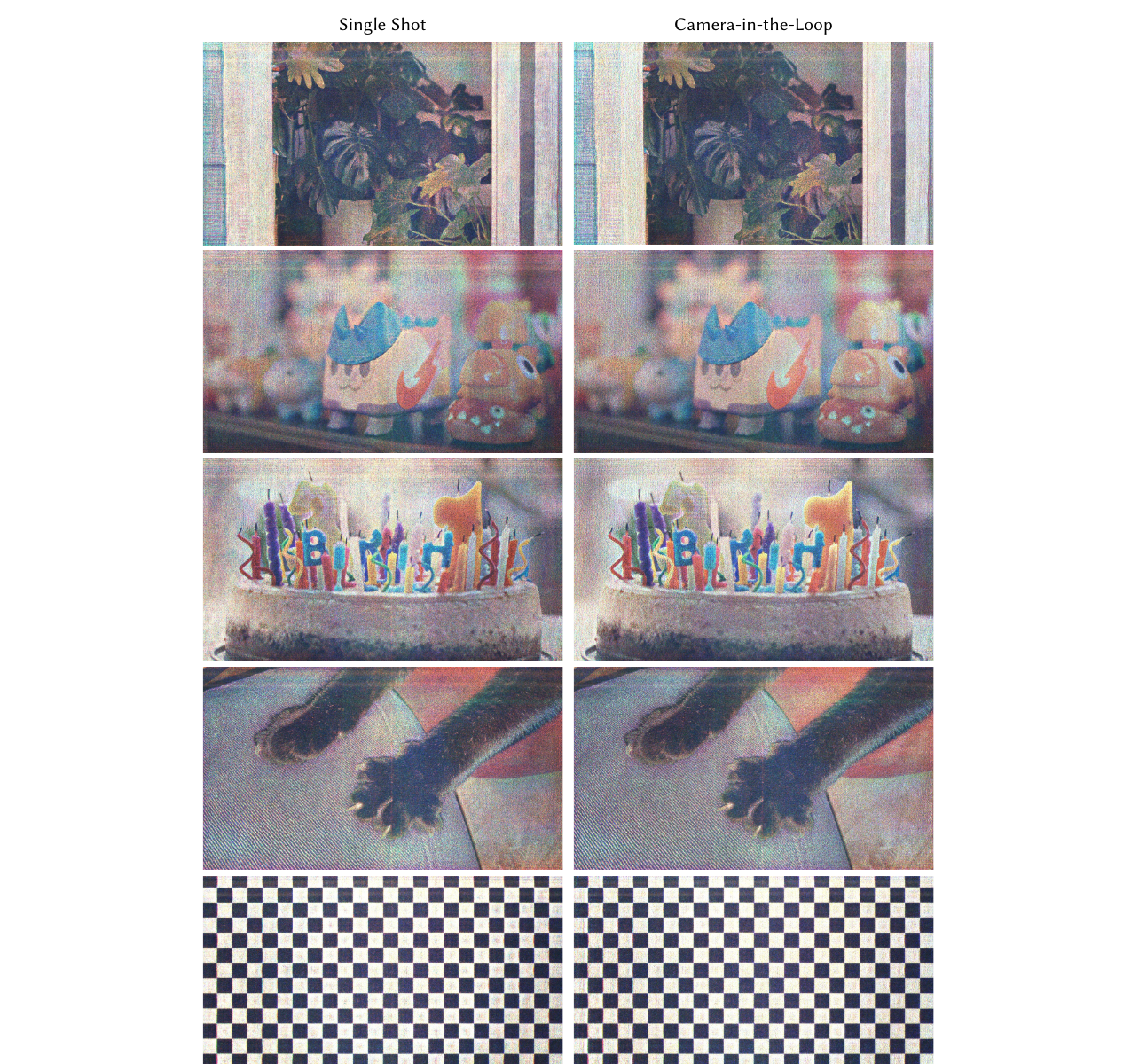}
    \caption{\textbf{Active camera-in-the-Loop (CiTL) reduces noise and improves color fidelity.} The first column of this image depicts experimentally captured color holograms.  The second columns shows images that were iteratively improved with a camera in the system using the active CiTL algorithm of ~\citet{Peng2020NeuralTraining}.
    }  
    \label{fig:CiTL}
\end{figure*}

\begin{figure*}
    \centering
    \includegraphics[clip, trim = 0in 9.275in 0in 0in, width=\textwidth]{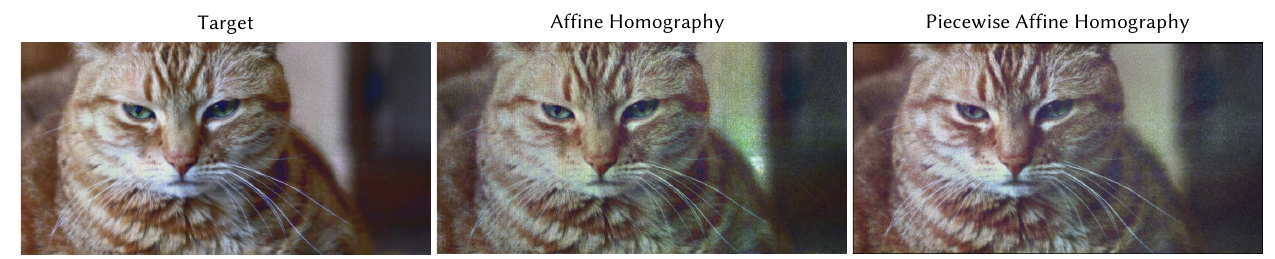}
    \caption{\textbf{Piecewise affine homography improves color fidelity for active CiTL.} The first column shows the target image. The second column shows the experimentally captured hologram optimized using active CiTL with a global affine homography. The third column depicts active CiTL with a piecewise affine homography, which reduces color artifacts and noise due to better alignment during optimization. Cat source image by Chris Erwin (CC-BY-2.0).
    }  
    \label{fig:PWA}
\end{figure*}

First, an SLM pattern is optimized using our learned simulation model and the computational graph is retained. This SLM pattern is then displayed and the resulting hologram is captured. A homography is applied to the captured hologram for each color channel to map it from camera space to simulation space. Our perceptual loss function is applied to the remapped captured hologram and target image. Backpropagation is performed using a computational graph saved during the forward pass, but the experimentally captured hologram is used in the loss function (instead of the simulated model output). This is the first time to our knowledge that active CiTL has been combined with a deep component to the forward model. Figure \ref{fig:CiTL} shows reduced noise and improved color fidelity for holograms generated with active CiTL. Since active CiTL uses the difference between the experimental capture and the target, the alignment between the two must be precise. We find that improved alignment using a piecewise affine homography, rather than a global affine homography, dramatically improves color fidelity. A comparison of this case is shown in Figure \ref{fig:PWA}.

\clearpage
\section{Additional Experimental Results and Failure Cases}

\begin{figure*}[!b]
    \centering
    \includegraphics[clip, trim = 0in 2.875in 0in 0in, width=0.85\textwidth]{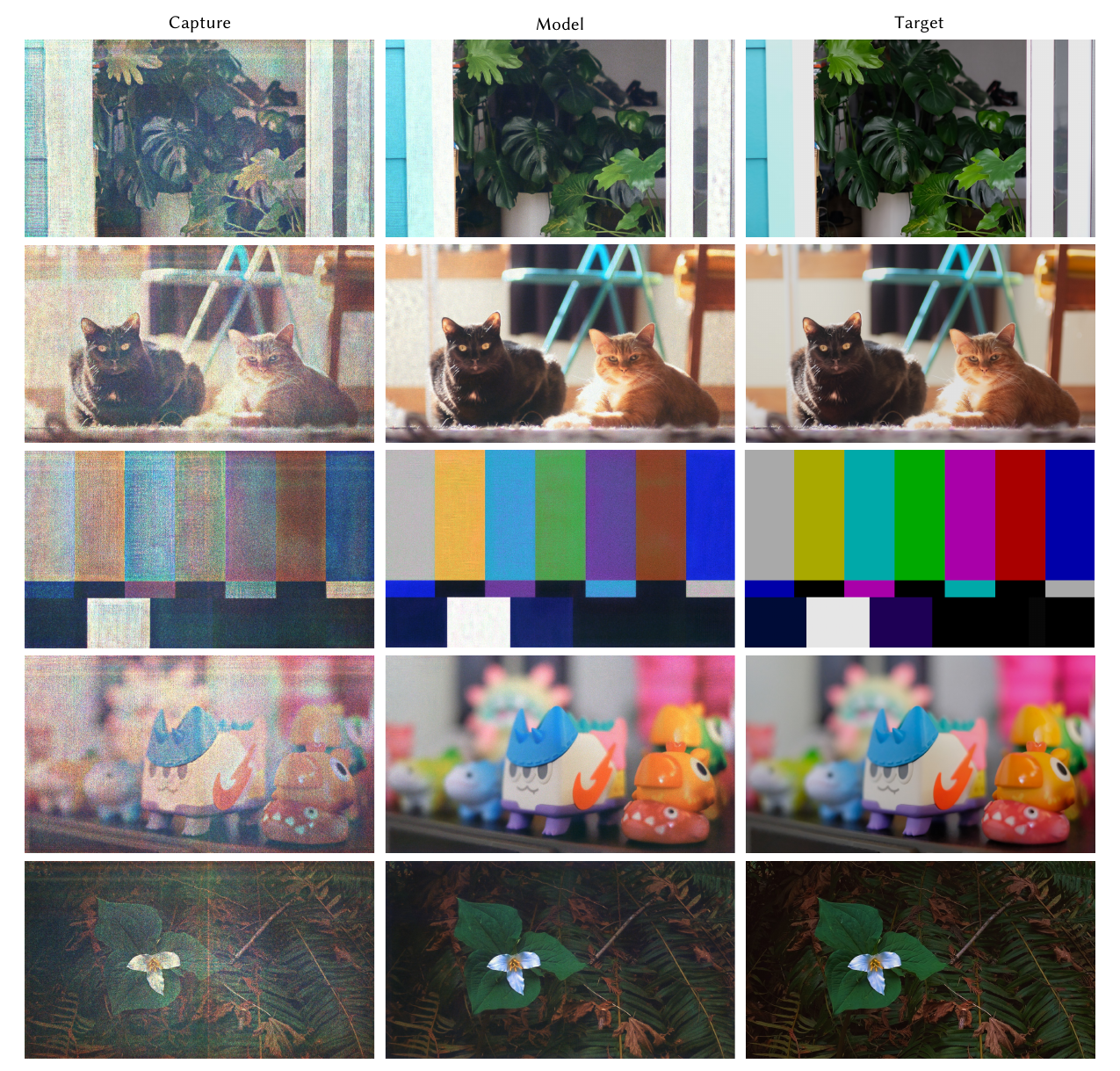}
    \caption{ \textbf{Additional simultaneous color holograms captured in experiment.}  The first column depicts holograms captured in experiment.  The second column shows the simulation output.  The third column depicts the target image. Although our system performs well on most natural scenes, unnatural images such as the bars in the center row are more challenging for our algorithm.
    }  
    \label{fig:AddResults}
\end{figure*}

\begin{figure*}[!b]
    \centering
    \includegraphics[clip, trim = 0in 4in 0in 0in, width=0.85\textwidth]{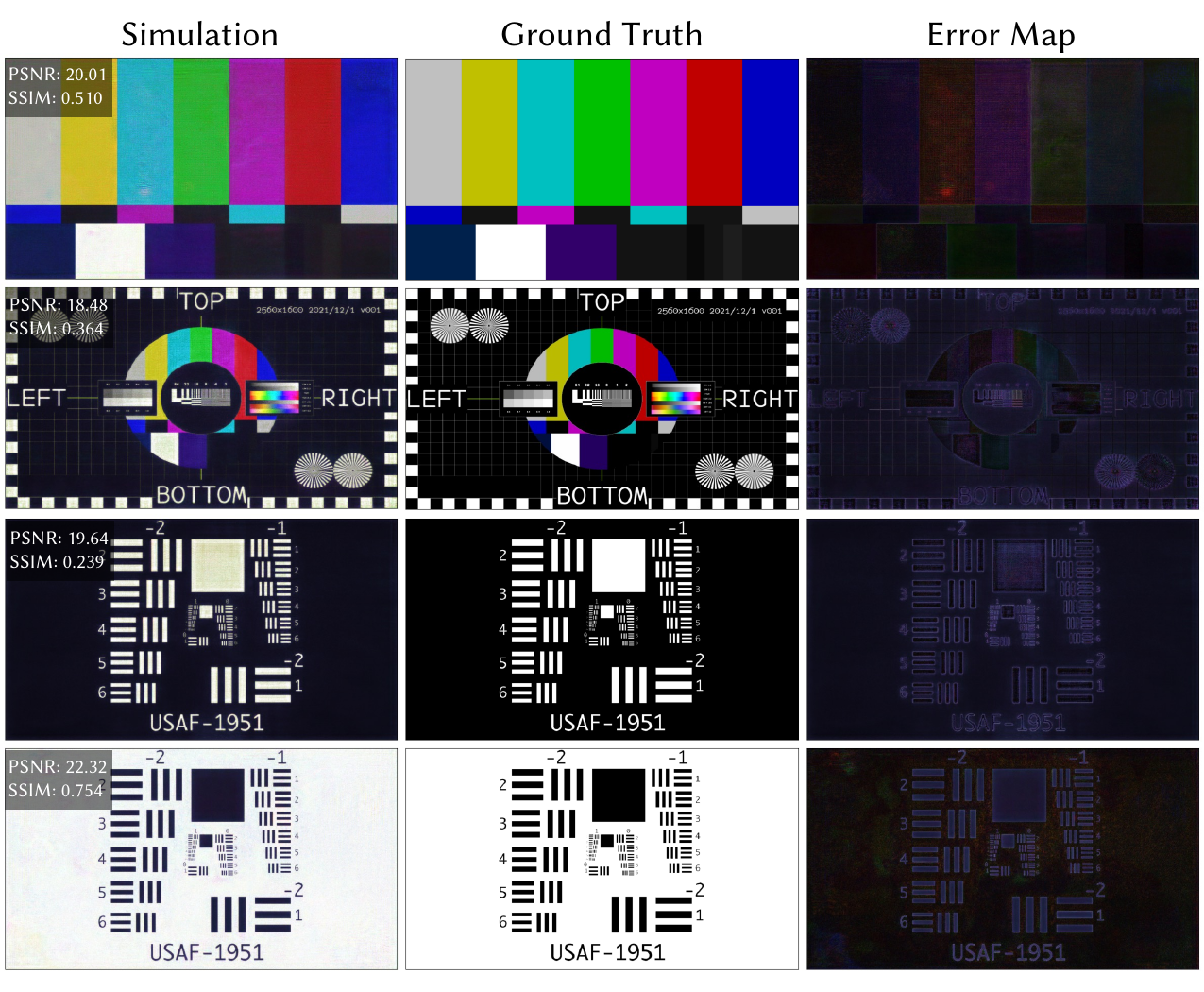}
    \caption{ \revised{\textbf{Failure case simulation results.}  The first column depicts holograms simulated using our proposed method with no U-Net component.  The second column shows the target image.  The an error map between the simulation and the ground truth. Our proposed method struggles to produce holograms that contain large patches of highly saturated colors.  Our method also struggles to produces white content on a black background as demonstrated in row three.
    }  }
    \label{fig:Failure}
\end{figure*}

Figure \ref{fig:AddResults} depicts additional captured results, which are intended to showcase a wider variety of scenes and include failure cases of our method. Our method has the most difficulty when the target has large, flat areas (i.e. textureless) of saturated color. Textureless targets lack high frequency information that can be leveraged by our loss function, leading to substantial artifacts such as color non-uniformity and ringing . These artifacts are particularly apparent in the image of colored bars in Fig. \ref{fig:AddResults}. Highly saturated images or ``unnatural'' images (like the colored bars) often fail due to disparate color channels, resulting in a single SLM pattern having to produce three holograms at the same plane with substantially different structures.  In contrast, natural images typically have similarly structured color channels.  \revised{We provide further simulation results of this failure case in Figure ~\ref{fig:Failure}.} 

\clearpage
\section{Additional Perceptual Loss Function Details}

\begin{figure}
    \centering
    \includegraphics[clip, trim = 0in 9.25in 0in 0in, width=\textwidth]{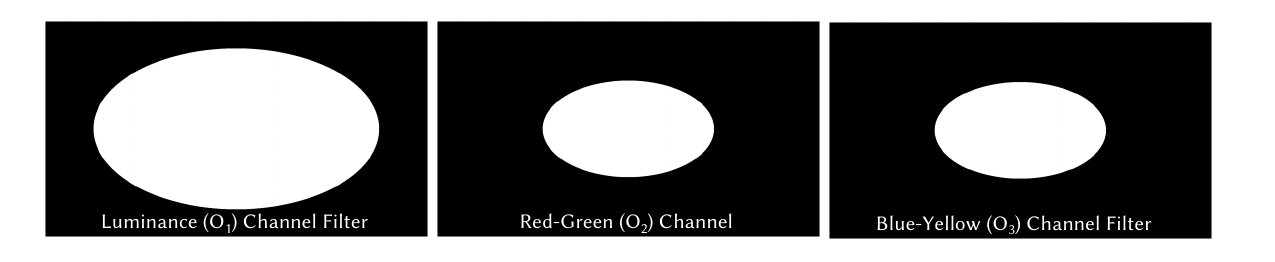}
    \caption{\textbf{Perceptual loss function filters in Fourier opponent color space.} The white areas of the filters pictured represents the pass band of the filter.  The luminance channel has a filter width of 75\% of Fourier space.  Both chrominance channels (Red-Green, Blue-Yellow) have filter widths of 45\% of Fourier space.
    }  
    \label{fig:Filters}
\end{figure}

Figure \ref{fig:Filters} shows a visualization of the perceptual loss function filters used in our algorithm. These filter sizes were kept constant regardless of the scene being optimized. To test the effectiveness of our perceptual loss function, we applied it to a personally captured dataset of 294 images. For each target image, an SLM pattern was optimized using both the traditional RGB loss function and our perceptual loss function. The resulting hologram was then captured, and the perceptual filter was applied. The PSNR, SSIM, and NMSE were calculated for the filtered simulated holograms and the perceptually filtered target image. The average metrics over the entire dataset are provided in Table \ref{loss_metrics}.

To make our system more general, we defined the perceptual loss filters relative to the maximum spatial frequency of the SLM instead of in physical units. However, by choosing the focal length of the eye piece, we can relate the filter sizes to physical quantities through the following relationship:
\begin{align}
\text{cutoff (cycles/deg)} = \frac{f}{2p}\cdot\frac{\pi}{180}\cdot \gamma
\end{align}
where $f$ is the focal length of the eye piece, $p$ is the SLM pixel size, and $\gamma$ is the filter width defined as a fraction of the Fourier space extent. In our simulations, $\gamma = 0.75$ for luminance and 0.45 for chrominance as shown in Fig.~\ref{fig:Filters}. Therefore, with an $\SI{8}{\micro\metre}$ pixel pitch and eye piece focal length of $\SI{36.6}{\milli\metre}$, the cutoffs correspond to 30 cycles/deg and 18 cycles/deg, for luminance and chrominance respectively. This is similar to perceptual measurements of human vision: \citet{mullen1985contrast} measure approximately 30 cycles/deg in luminance and 11-12 cycles/deg in chrominance, which is actual slightly less accuity in color than we target in our images.

\begin{table}[ht]
    \centering
        \begin{tabular}[t]{lccc}
        & PSNR & SSIM & NMSE \\
        \hline
        RGB Loss Function & 20.11 & 0.603 & 0.010  \\
        \hline
        Perceptual Loss Function & 26.58 & 0.869 & 0.003  \\
        \hline
        \end{tabular}
        \vspace{3mm}
        \caption{ A comparison of the average PSNR, SSIM, and NMSE for holograms optimized with the traditional RGB loss function and perceptual loss function.  The metrics were calculated between the perceptually filtered simulated holograms and the perceptually filtered target.  The data set used was a personally captured set of 294 images of natural scenes.}
        \label{loss_metrics}
\end{table}

\clearpage
\section{The Effect of Bit Depth on Hologram Quality}
\label{sec:Bits}
\begin{figure*}[!b]
    \centering
    \includegraphics[clip, trim = 0in 2.875in 0in 0in, width=0.8\textwidth]{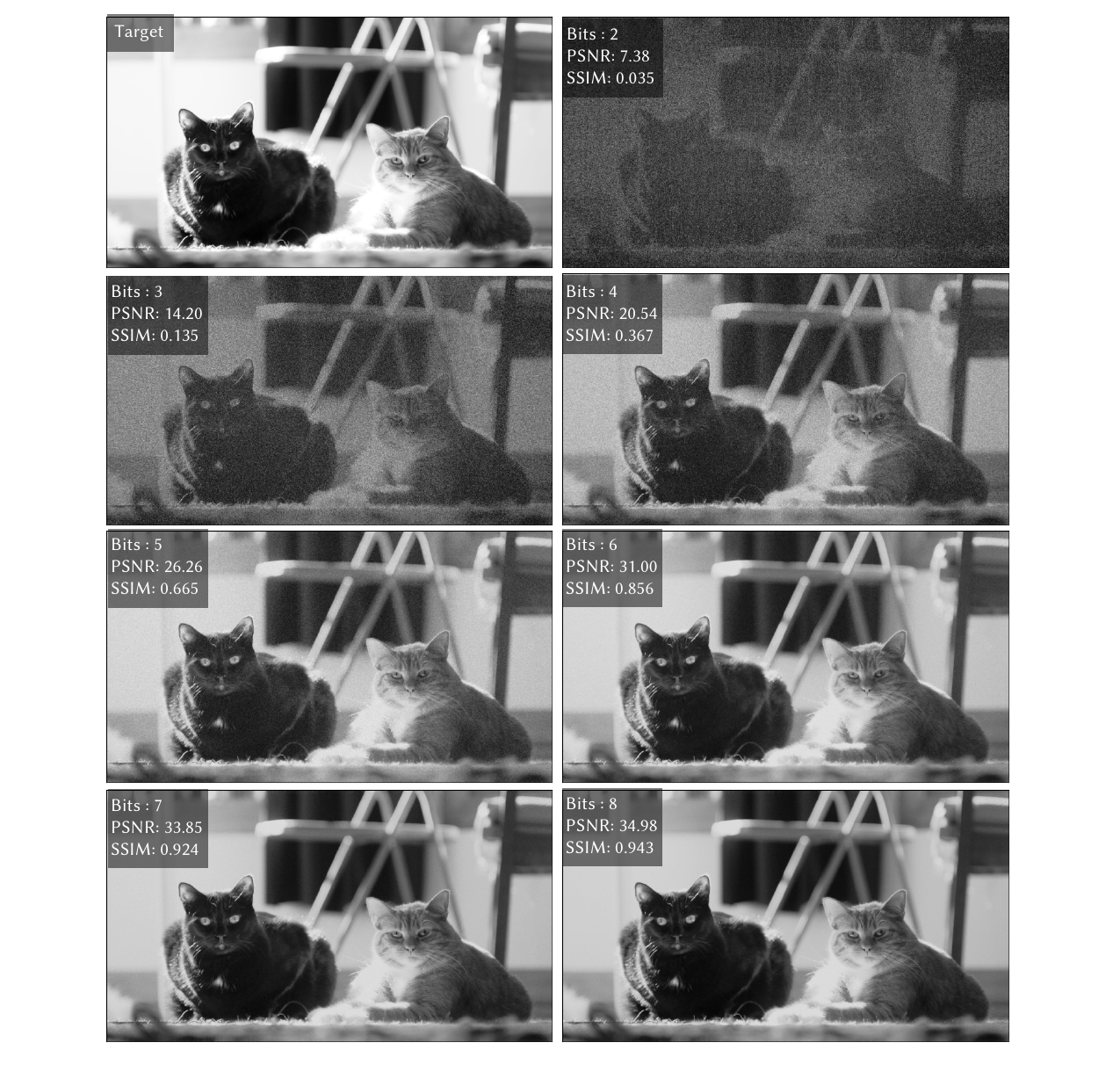}
    \caption{ \textbf{An analysis of SLM bit depth on hologram quality in simulation}  We simulate holograms using SLMs of 2 to 8 bits.  The target image is pictured in the top left of the figure.  One should note the rapidly increasing drop off in both PSNR and SSIM between 5 and 6 bits.
    }  
    \label{fig:Bits}
\end{figure*}

\begin{figure*}[!b]
    \centering
    \includegraphics[clip, trim = 0in 1.675in 0in 0in, width=1\textwidth]{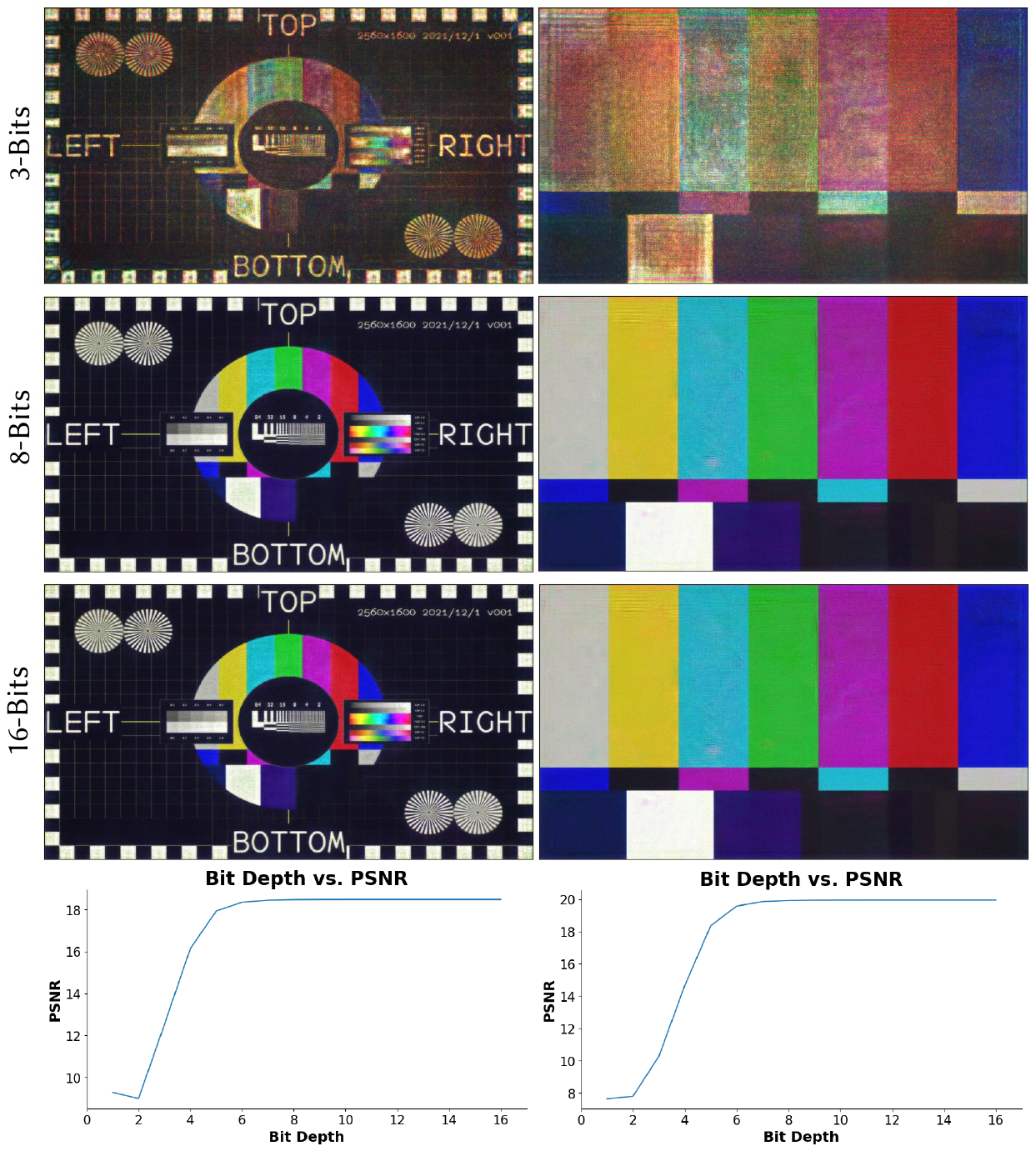}
    \caption{ \revised{\textbf{An analysis of SLM bit depth on simultaneous color hologram quality in simulation}  We simulate holograms using SLMs of 1 to 16 bits.  Rows 1-3 depict the simulated simultaneous color holograms with an SLM of bit depth 3, 8, and 16 respectively.  Row 4 depicts the bit depth vs. PSNR.  PSNR is calculated with the perceptually filtered ground truth and perceptually filtered simulated hologram. 
    } } 
    \label{fig:SimulBits}
\end{figure*}

The effect of quantization on hologram quality is an important consideration when choosing an extended phase SLM. We define the effective bit depth as the number of bits contained in a $2\pi$ interval of the extended range. For example, the effective bit depth of an 8-bit SLM with a phase range of $8\pi$ is 6 bits as each $2\pi$ interval contains 64 discrete samples i.e. 6 bits. To determine the minimum bit depth required for adequate image quality, we simulated holograms using an SLM with a $2\pi$ phase range and bit depths from 2 bits to 8-bits. Simulations are done by optimizing the hologram with gradient descent, then quantizing to the target bit depth. A significant drop off in both PSNR and SSIM was observed between 5 and 6 bits, as depicted in Fig. \ref{fig:Bits}. This suggests that the minimum effective bit depth required for an extended phase SLM is 6 bits. Since most commercially available SLMs are 8 bits, this suggests that the maximum phase range in any channel should be $8\pi$, which aligns well with the SLM used in our experiments (maximum phase range of $7.4\pi$ in the blue channel).  \revised{ Additionally, we find that no image quality improvement is achieved for simultaneous color holograms once each color channel has a bit depth of at least 6 bits across a $2\pi$ phase range.  The results of this simulation are displayed in Figure ~\ref{fig:SimulBits}.}

\clearpage
\section{The Effect of Phase Range on Simultaneous Color Hologram Quality}The phase range of an SLM plays a critical role in the arena of simultaneous color holography. In this context, we must ensure that the phase range for each of the three color channels is sufficiently different. This differentiation is crucial as it allows for the production of unique holograms in each color channel, overcoming the inherent depth-wavelength ambiguity associated with the ASM kernel. However, there's a trade-off. An SLM with a larger phase range can potentially degrade the image quality by reducing the effective bit depth, as detailed in Section \ref{sec:Bits}. Additionally, SLMs with larger phase ranges tend to have slower refresh rates, counterbalancing some of the advantages gained from simultaneous illumination.

\begin{figure*}[!b]
    \centering
    \includegraphics[clip, trim = 0in 1in 3.8in 0in, width=0.5\textwidth]{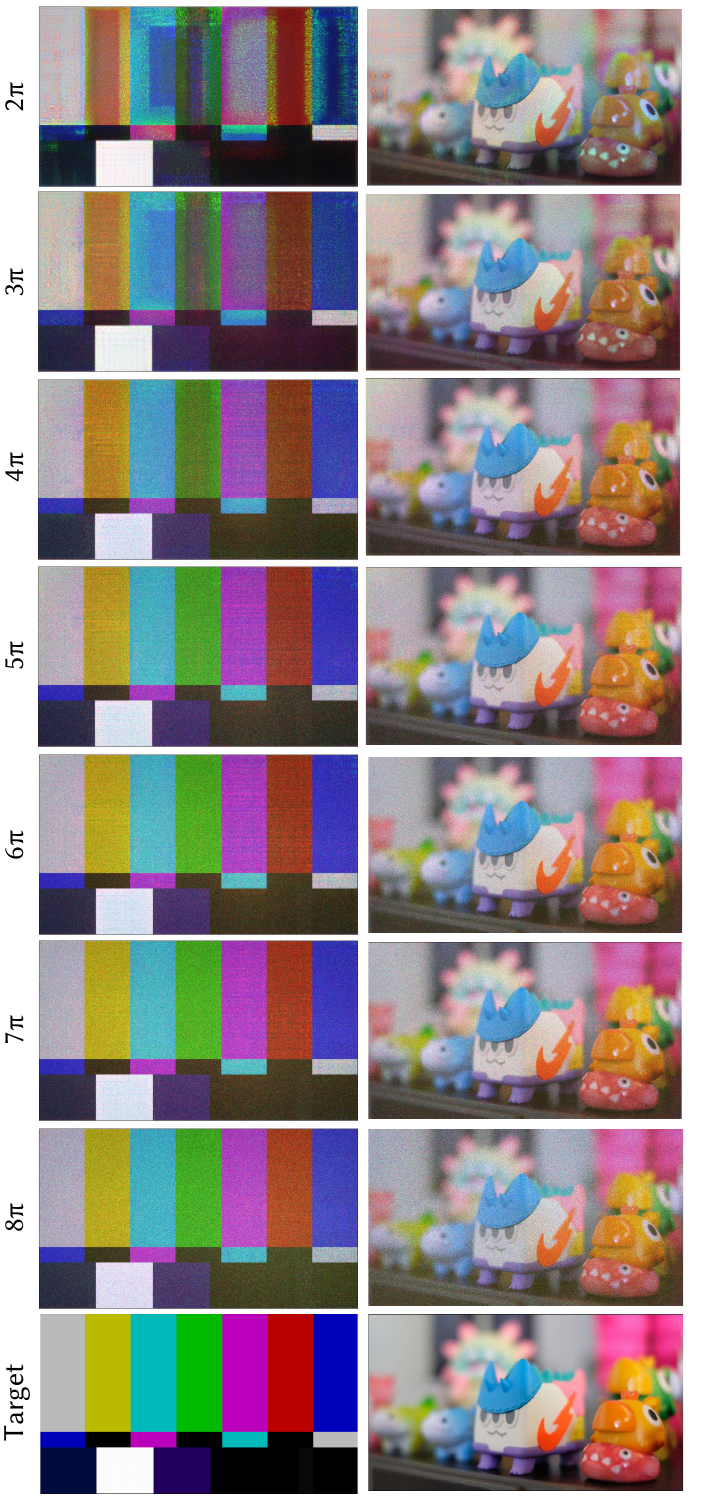}
    \caption{ \textbf{A demonstration of SLM phase range on hologram quality in simulation}] The images in the first column represent unnatural images, while the second column shows natural images. Each row corresponds to a specific maximum phase range in the blue channel, denoted by the labels of pi. The phase range in the red channel is fixed at 2$pi$, and the green channel's phase range is the average of the red and blue channels. Our findings demonstrate that a minimum phase range of 5$pi$ for unnatural images and 4$pi$ for natural images in the blue channel produces satisfactory holographic reconstructions. This is anticipated given that unnatural images typically have more unique color channels, thus necessitating a larger phase range, while natural images with their similar color channels require less phase variation.
    }  
    \label{fig:Phase}
\end{figure*}

Thus, when choosing an SLM for this purpose, the aim should be to select one that offers the minimal phase range necessary to achieve the desired image quality.  In our study depicted in Figure \ref{fig:Phase}, we conducted a series of experiments with various phase values to find a range that delivers adequate quality simultaneous color holograms for both natural and unnatural images. We utilized gradient descent with a naive ASM forward model for each SLM pattern. The L2 norm between the target image and the simulated hologram served as the loss function.

Our experiment involved fixing the red channel phase range at 2$pi$ and incrementally increasing the blue channel phase range from 2$pi$ to 8$pi$. We set the green channel's phase range as the average of the red and blue channels. Our findings reveal that for unnatural images, a minimum phase range of 5$pi$ in the blue channel is sufficient, and for natural images, a phase difference of 4$pi$ is adequate.  We find this result unsurprising as unnatural images have largely unique color channels while natural images have fairly similar color channels.  This suggests the task of optimizing an SLM pattern for natural images is easier and consequently requires less phase.

\clearpage
\section{Bit and Depth Division Implementation Details and Analysis}
In this section we provide our implementation details of bit and depth division holography.  Additionally, we analyze the methods for SLMs of various phase ranges.  We implement bit division largely as laid out by \citet{Jesacher2014ColourRange}.  First we calculated the three color channels SLM patterns using a modified Gerchberg-Saxton approach assuming a $2\pi$ phase range in each color channel.  Instead of using the Fourier transform for propagation as in \citet{Jesacher2014ColourRange}, we use ASM match our other results.  This is run until convergence, and 3 unique SLM patterns are produced.  These SLM patterns are then combined via an optimization problem as described by \cite{Jesacher2014ColourRange}. We then used the combined SLM pattern to simulate a color hologram at the sensor plane.  Pseudocode for this algorithm is provided by Algorithm \ref{alg:bit_division_holography}.

\begin{algorithm}[htb]
\caption{Bit Division Holography}
\label{alg:bit_division_holography}
\begin{algorithmic}[1]
\State \textbf{Step 1: Individual color channel SLM pattern calculation}
    \For{each color channel (Red, Green, Blue)}
    \State Get the target intensity for the current color channel: $I_{\text{target}}$
    \State Initialize the hologram phase pattern for the current color channel: $H_0$
    \State $A = Amplitude(Source) * e^{iH_0}$
    \While{not converged}
      \State Perform the modified Gerchberg-Saxton algorithm for the current color channel:
        \State $B = angularSpectrumPropagation(A, d_{color}, \lambda_{color})$
        \State $C = Amplitude(I_{\text{target}}) * e^{iPhase(B)}$
        \State $D = angularSpectrumPropagation(C, d_{color}, \lambda_{color})$
        \State $A = Amplitude(Source) * e^{iPhase(D)}$   
    \EndWhile
    
    \State Store the optimized hologram for the current color channel: $H_{color}$
    \EndFor

\State \textbf{Step 2: Hologram combination}
    \State Combine the three optimized holograms by solving the optimization problem described by Jesacher et al. [2014]
    
\State \textbf{Step 3: Hologram simulation}
    \State Simulate a color hologram at the sensor plane using the combined hologram

\State \textbf{Output} the final optimized holograms
\end{algorithmic}
\end{algorithm}

We choose to implement the depth division method using gradient descent-based optimization rather than a modified Gerchberg-Saxton (GS) algorithm for multiplane holograms originally proposed by \citet{Makowski2008ColorfulHologram, Makowski2010ColorHolograms} for depth division holography. Since we use gradient descent in our approach, we determined this was a more fair comparison. In our implementation the SLM pattern is first converted to a complex field.  The complex field is then propagated to $depth =\SI{68}{\milli\metre}, \SI{80}{\milli\metre}, \SI{100}{\milli\metre}$ using the ASM kernel for the red color channel. These correspond to the replica planes. The intensity of the of the fields are then calculated at each target plane and compared to the blue, green, and red channels, respectively, using an L2 loss function.  Backpropagation is then used to calculate the gradients of the loss function with respect the SLM voltage values and then update these voltages.
Pseudocode for this algorithm is provided by Algorithm \ref{alg:depth_division_holography}.

\begin{algorithm}[htb]
\caption{Depth Division Holography}
\label{alg:depth_division_holography}
\begin{algorithmic}[1]
\State \textbf{Initialization}:
\State - Choose a depth to display the multicolor holgoram: $d_{target}$

\State \textbf{Step 1: Calculate depth planes}
    \State Set the red channel propagation distance, $d_{red}$, equal to the target depth: $d_{red}=d_{target}$
    \State Determine the ratio r of the propagation distance to the wavelength for the red color: $r = d_{red} / \lambda_{red}$
    \State Apply this ratio to the blue and green wavelengths to calculate the propagation distances (depth planes) where their angular spectrum kernel equals that of the red color: $d_{blue} = r * \lambda_{blue}$, $d_{green} = r * \lambda_{green}$

\State \textbf{Step 2: Initialize multiplane hologram optimization}
    \State Initialize parameters for gradient descent optimization, including learning rate, number of iterations, and initial hologram

\State \textbf{Step 3: Gradient descent optimization}
    \For{iteration from 1 to numIterations}
    \State Initialize target intensities for red, green, and blue
    \For{d in $[d_{red}, d_{green},d_{blue}]$}
    \State Propagate the field using angularSpectrumPropagation: $propagatedField = angularSpectrumPropagation(currentHologram, d, \lambda_{red})$
    \State Compute targetIntensity from propagatedField
    \State Append targetIntensity to respective color targetIntensities
    \EndFor
    
    \State Compute the loss: $loss = computeLoss(target, targetIntensities)$
    \State \textbf{print}("Iteration:", iteration, "Loss:", loss)
    
    \State Compute the gradient: $gradient = computeGradient(target, targetIntensities, currentHologram)$
    \State Update currentHologram: $currentHologram = currentHologram - learningRate * gradient$
    \EndFor

\State \textbf{Step 4: Return the optimized multiplane hologram}
    \State \textbf{Output} the final optimized hologram: currentHologram
\end{algorithmic}
\end{algorithm}

\begin{figure*}
    \centering
    \includegraphics[clip, trim = 0in 4.575in 0in 0in, width=\textwidth]{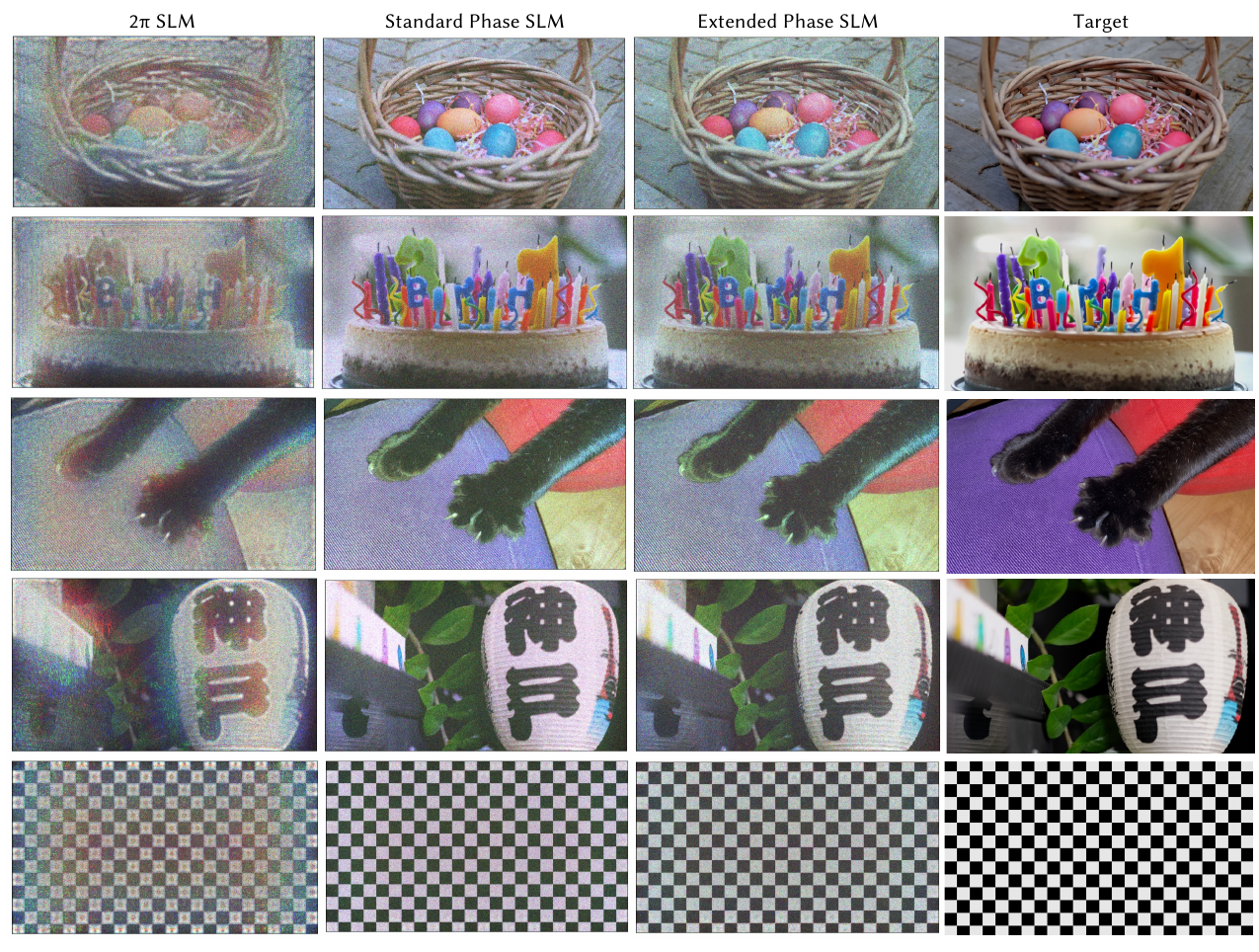}
    \caption{ \textbf{SLM phase range affects hologram quality for bit division holography.} Bit division takes advantage of the extended phase range of the SLM, so does not perform well with an SLM with only $2\pi$ phase range per channel (left column). With a `` standard'' SLM with realistic wavelength dependence to the phase, bit division performs better. It works best with the extended phase range of the simulated Holoeye Pluto that we use for our experiments.}  
    \label{fig:ExtenedBits}
\end{figure*}

\begin{figure*}
    \centering
    \includegraphics[clip, trim = 0in 4.575in 0in 0in, width=\textwidth]{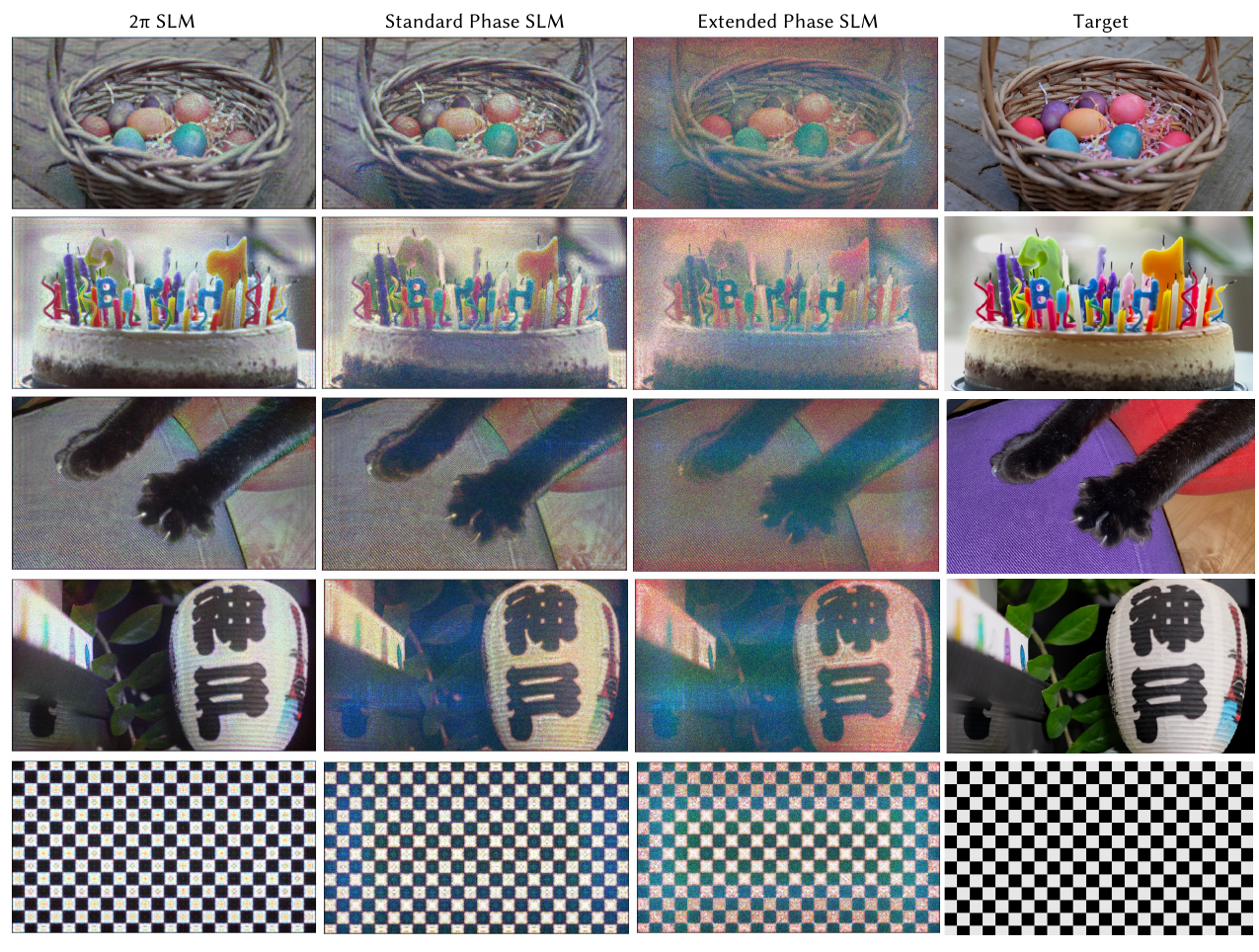}
    \caption{ \textbf{SLM phase range affects hologram quality for depth division holography.} The depth division approach assumes no wavelength dependence of the SLM, which is simulated in the first column. With a standard SLM with $2\pi$ phase in red and realistic wavelength dependence (second column) the results are slightly degraded due to the violation of the no wavelength dependence assumption. Finally, with the extended phase range of the simulated Holoeye Pluto SLM, the results show significant color artifacts and noise.}  
    \label{fig:ExtenedDepth}
\end{figure*}

We implement both the bit and depth division holography methods for 3 simulated SLMs.  The first SLM has a uniform $2\pi$ phase range in each color channel.  This phase range is optimal for depth division, but performs the worst of the simulated SLMs for bit division, demonstrating how bit division relies on extended SLM phase range. The next simulated SLM is an arbitrary standard SLM i.e. not extended phase.  We model this SLM to have $2\pi$ phase range in red, $2.7\pi$ in green, and $3.4\pi$ in blue.  The simulated holograms increase in quality from the $2\pi$ SLM for the bit division method, but decrease in quality for depth division.  Finally we simulate the Holoeye Pluto SLM used in our experimental setup.  This SLM has a $2.4\pi$ phase range in red, $5.9\pi$ phase range in green, and $7.4\pi$ phase range in blue.  The results for depth division continue to degrade with this SLM, since the depth division algorithm does not take into account the wavelength-dependent response of the SLM. The results improve for bit division with the additional extended phase.  This suggests that that phase diversity across channels provides the best performance for bit division holography, while phase uniformity across channels provides the best performance for depth division holography.  The results of the outlined experiment can be found in Figs. \ref{fig:ExtenedBits} and \ref{fig:ExtenedDepth}.

\begin{figure*}
    \centering
    \includegraphics[clip, trim = 0in 2.5in 0in 0in, width=0.9\textwidth]{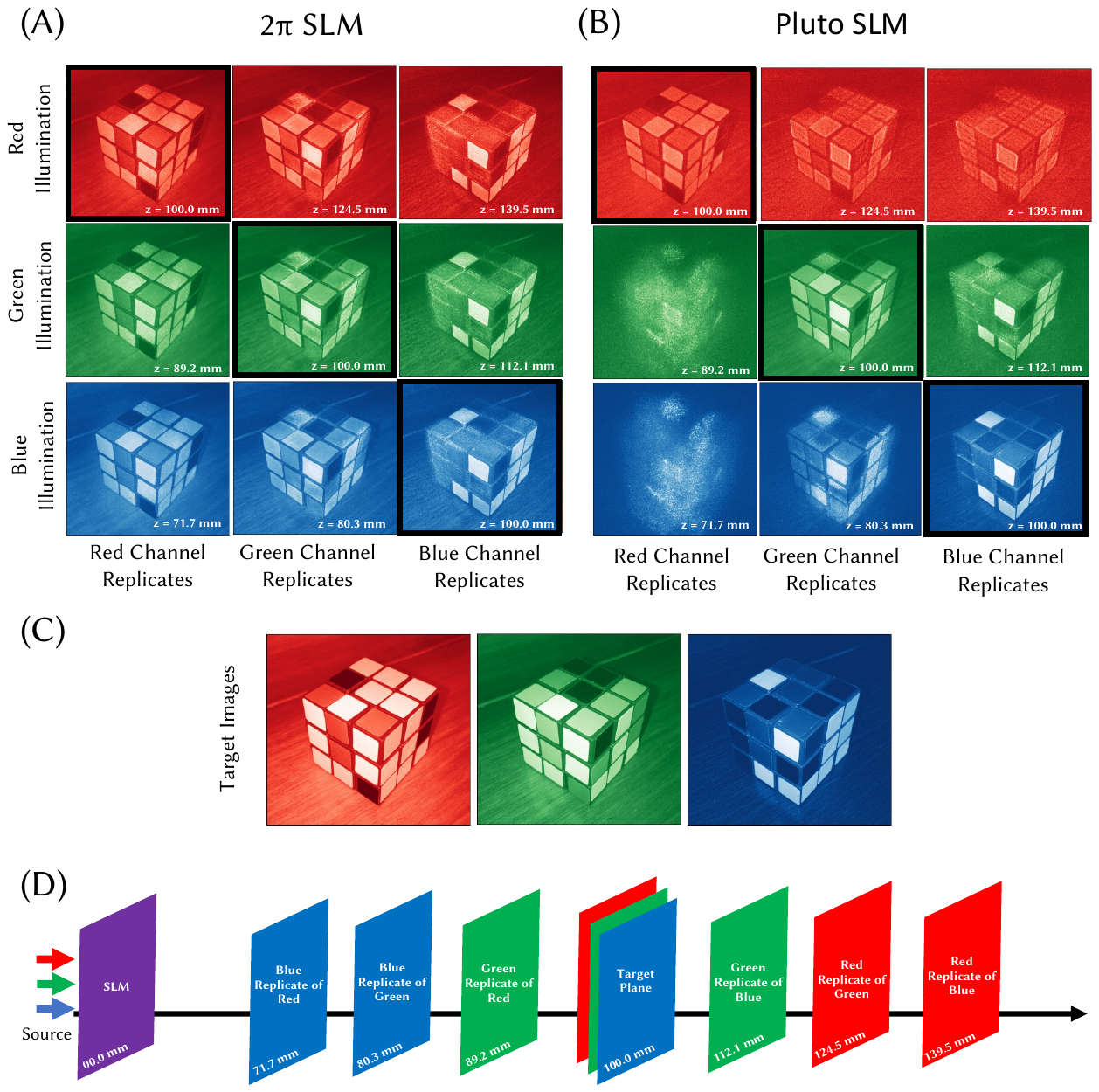}
    \caption{\textbf{Extended phase range reduces depth replicas in simulation.}
    (A) Using an SLM with a uniform $2\pi$ phase range across all channels leads to strong depth replicas, which reduce image quality at the target plane (diagonal) compared to the targets (C) and add in-focous content at depths that should be defocused. By using the extended phase Holoeye Pluto-2.1-Vis-016 SLM (with  Red: $2.4\pi$, Green: $5.9\pi$, Blue: $7.4\pi$ phase ranges), depth replicas are significantly reduced (B), improving the quality of target plane holograms and creating defocused content at other depths. (D) Schematic illustrating the positions of the replicate planes and target plane. Rubik's cube source image by Iwan Gabovitch (CC BY 2.0).}
    \label{fig:Replicates}
\end{figure*}

\clearpage

% \appendix

% \section{Appendix}
\end{document}